\documentclass{article}
\usepackage{amsmath,amsfonts,amsthm,amssymb}

\usepackage[final]{nips_arxiv}

\usepackage[utf8]{inputenc} 
\usepackage[T1]{fontenc}    
\usepackage{hyperref}       
\usepackage{url}            
\usepackage{booktabs}       
\usepackage{amsfonts}       
\usepackage{nicefrac}       
\usepackage{microtype}      
\usepackage{amsmath}
\usepackage{amssymb}
\usepackage{graphicx}
\usepackage{ifthen}
\usepackage[utf8]{inputenc}
\usepackage{color}
\usepackage{algorithm,algorithmicx,algcompatible,algpseudocode}
\usepackage{framed}
\usepackage{enumitem}
\usepackage{xparse}
\DeclareDocumentCommand{\citetm}{ O{} O{} m }{\citet[#1][#2]{#3}}
\DeclareDocumentCommand{\citeta}{ O{} O{} m }{\citet[#1][#2]{#3}}
\DeclareDocumentCommand{\citepm}{ O{} O{} m }{\citep[#1][#2]{#3}}
\DeclareDocumentCommand{\citepa}{ O{} O{} m }{\citep[#1][#2]{#3}}

\usepackage{blkarray, bigstrut}
\makeatletter
\newbox\BA@first@box
\makeatother
\usepackage{multirow,varwidth}
\newcommand\topstrut[1][1.2ex]{\setlength\bigstrutjot{#1}{\bigstrut[t]}}
\newcommand\botstrut[1][0.9ex]{\setlength\bigstrutjot{#1}{\bigstrut[b]}}

\newcommand\sC{\ensuremath{\mathcal{C}}}

\newcommand\sP{\ensuremath{\mathcal{P}}}

\newcommand\bE{\ensuremath{\mathbf{E}}}

\newcommand\bI{\ensuremath{\mathbf{I}}}

\newcommand\bP{\ensuremath{\mathbf{P}}}

\newcommand\bR{\ensuremath{\mathbf{R}}}

\newcommand\Bx{\ensuremath{\mathbb{x}}}
\newcommand\By{\ensuremath{\mathbb{y}}}

\newcommand\p[1]{\ensuremath{\left( #1 \right)}} 

\newcommand\eqdef{\ensuremath{\stackrel{\rm def}{=}}} 

\ifthenelse{\isundefined{\definition}}{\newtheorem{definition}{Definition}}{}
\ifthenelse{\isundefined{\assumption}}{\newtheorem{assumption}{Assumption}}{}
\ifthenelse{\isundefined{\hypothesis}}{\newtheorem{hypothesis}{Hypothesis}}{}
\ifthenelse{\isundefined{\proposition}}{\newtheorem{proposition}{Proposition}}{}
\ifthenelse{\isundefined{\theorem}}{\newtheorem{theorem}{Theorem}}{}
\ifthenelse{\isundefined{\lemma}}{\newtheorem{lemma}{Lemma}}{}
\ifthenelse{\isundefined{\corollary}}{\newtheorem{corollary}{Corollary}}{}
\ifthenelse{\isundefined{\alg}}{\newtheorem{alg}{Algorithm}}{}
\ifthenelse{\isundefined{\example}}{\newtheorem{example}{Example}}{}
\newcommand{\E}{\ensuremath{\mathbb{E}}} 

\def\[#1\]{\begin{align}#1\end{align}}
\def\(#1\){\begin{align*}#1\end{align*}}

\newcommand{\bprf}{\begin{proof}}
\newcommand{\eprf}{\end{proof}}
\newcommand{\blem}{\begin{lemma}}
\newcommand{\elem}{\end{lemma}}
\newcommand{\eps}{\epsilon}

\DeclareMathOperator{\Uniform}{Uniform}

\newcommand{\oo}{\mathcal{O}}

\newcommand{\op}{\mathrm{op}}

\DeclareMathOperator{\Var}{Var}

\definecolor{mydarkblue}{rgb}{0,0.08,0.45}
\hypersetup{ %
  colorlinks=true,
  linkcolor=mydarkblue,
  citecolor=mydarkblue,
  filecolor=mydarkblue,
  urlcolor=mydarkblue}
\setcitestyle{authoryear,round,citesep={;},aysep={,},yysep={;}}
\usepackage{tikz}
\usetikzlibrary{patterns,positioning,fit,calc,intersections}

\newcommand{\todo}[1]{{\color{red} [TODO: {#1}]}}

\newcommand{\Obs}{\mathcal{E}}

\newcommand{\good}{\sC}
\newcommand{\goodapprox}{\sC'}

\newcommand{\bad}{\overline{\sC}}
\newcommand{\goodfrac}{\alpha}
\newcommand{\goodraters}{\alpha}

\newcommand{\error}{\epsilon}
\newcommand{\failprob}{\delta}
\newcommand{\Aavg}{A^*}
\newcommand{\Mavg}{M^*}
\newcommand{\Aobs}{\tilde{A}}

\newcommand{\rtrue}{r^*}
\newcommand{\ravg}{r^*}

\newcommand{\robs}{\tilde{r}}

\DeclareMathOperator{\rank}{rank}
\newcommand{\M}{\tilde{M}}
\newcommand{\Mm}{M^*}
\newcommand{\A}{\tilde{A}}
\newcommand{\Aa}{B}

\title{Avoiding Imposters and Delinquents: Adversarial Crowdsourcing and Peer Prediction}

\author{
  Jacob Steinhardt \\
  Computer Science Department \\
  Stanford University \\
  {\tt jsteinhardt@cs.stanford.edu}
\And
	Gregory Valiant \\
  Computer Science Department \\
  Stanford University \\
  {\tt valiant@stanford.edu}
\And
	Moses Charikar \\
  Computer Science Department \\
  Stanford University \\
  {\tt moses@cs.stanford.edu}
}

\begin{document}
\algrenewcomment[1]{\hfill $\triangleright$ #1}
\renewcommand{\paragraph}[1]{\textbf{#1}}

\maketitle

\begin{abstract}
We consider a crowdsourcing model in which $n$ workers are asked to 
rate the quality of $n$ items previously generated by other workers.
An unknown set of $\goodraters n$ workers generate reliable ratings, 
while the remaining workers may behave arbitrarily and possibly adversarially. 
The manager of the experiment can also manually evaluate the quality of a 
small number of items, and wishes to curate together almost all 
of the high-quality items with at most an $\error$ fraction of 
low-quality items.  
Perhaps surprisingly, we show that this is possible with an 
amount of work required of the manager, and each worker, that does not scale 
with $n$: the dataset can be curated with
$\tilde{\oo}\p{\frac{1}{\beta\goodraters^3\error^4}}$ 
ratings per worker, and $\tilde{\oo}\p{\frac{1}{\beta\error^2}}$ 
ratings by the manager, where $\beta$ is the fraction of 
high-quality items.
Our results extend to the more general setting of peer prediction, 
including peer grading in online classrooms.

\end{abstract}
\setlength{\textfloatsep}{8pt}

\section{Introduction}
\label{sec:intro}


How can we reliably obtain information from humans, given that the humans 
themselves are unreliable, and might even have incentives to mislead us?
Versions of this question arise in crowdsourcing \citepa{vuurens2011spam}, 
collaborative knowledge generation \citepm{priedhorsky2007creating}, peer grading 
in online classrooms \citepm{piech2013tuned,kulkarni2015peer},  aggregation 
of customer reviews \citepa{harmon2004amazon}, and the generation/curation of large datasets \citepm{deng2009imagenet}. A key challenge is to ensure 
high information quality despite the fact that many people interacting with 
the system may be unreliable or even adversarial.
This is particularly relevant when raters have an incentive to collude and 
cheat as in the setting of peer grading, as well as reviews on sites like 
Amazon and Yelp, 
where artists and firms are incentivized to manufacture positive reviews 
for their own products and negative reviews for their rivals \citepa{harmon2004amazon,mayzlin2012promotional}.

One approach to ensuring quality is to use \emph{gold sets} --- questions where 
the answer is known, which can be used to assess reliability on unknown questions. 
However, this is overly constraining --- it does not make sense for open-ended 
tasks such as knowledge generation on wikipedia, nor even for crowdsourcing 
tasks such as ``translate this paragraph'' or ``draw an interesting picture'' 
where there are different equally good answers.   This approach may also fail 
in settings, such as 
peer grading in massive online open courses, where 
students might collude to inflate their grades.

In this work, we consider the challenge of using crowdsourced human ratings to accurately and efficiently evaluate a large dataset of content.   In some settings, such as peer grading, the end goal is to obtain the accurate evaluation of each datum; in other settings, such as the curation of a large dataset, accurate evaluations could be leveraged to select a high-quality subset of a larger set of variable-quality (perhaps crowd-generated) data.  

There are several confounding difficulties that arise in extracting accurate evaluations.  First, many raters may be unreliable and give evaluations 
that are uncorrelated with the actual item quality; 
second, some reliable raters might be harsher or more lenient than others; 
third, some items may be harder to evaluate than others 
and so error rates could vary from item to item, even among reliable raters; 
finally, some raters may even collude or want to hack the system. 
This raises the question: can we obtain 
information from the reliable raters, without knowing who they are a priori?

In this work, we answer this question in the affirmative, under surprisingly 
weak assumptions:
\begin{itemize}[itemsep=2pt,topsep=0pt,parsep=0pt,partopsep=0pt,leftmargin=30pt]
\item We do not assume that there is a ``gold set'' or other cheap way to judge 
      worker performance; instead, we rely on a small number of our own (potentially noisy) post hoc judgments. 
\item We do not assume that the majority of workers are reliable.
\item We do not assume that the unreliable workers conform to any statistical 
      model; they could behave fully adversarially, in collusion with each other 
      and with full knowledge of how the reliable workers behave.
\item We do not assume that the reliable worker ratings match our own, but only that they are 
      ``approximately monotonic'' in our ratings, in a sense that will be 
      formalized later.
\end{itemize}
For concreteness, we describe a simple formalization of the crowdsourcing 
setting (our actual results hold in a more general setting). 
There are $n$ raters and $n$ items to evaluate, which have an unknown 
quality level in $[0,1]$. At least $\alpha n$ workers are ``reliable'' in that 
their judgments match our own in expectation, and they make independent errors.
We assign each worker to evaluate at most $k$ randomly selected items. 
In addition, we ourselves judge $k_0$ items. Our goal is to 
recover the \emph{$\beta$-quantile}: the set $T^*$ of the $\beta n$ highest-quality items. 
Our main result is the following:

\begin{figure}[b]
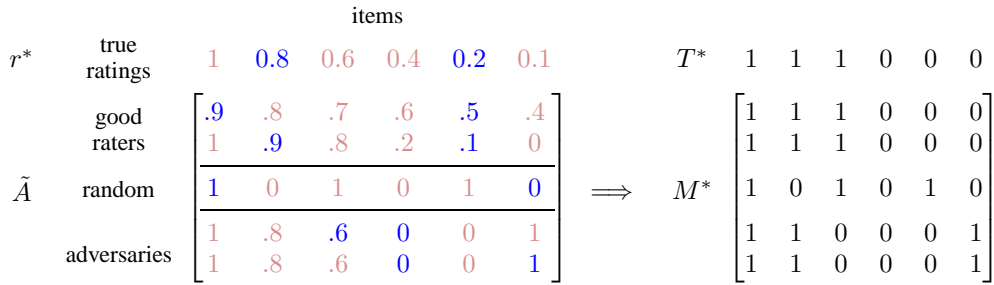

\centering
\FrameSep0pt
\newcommand{\ob}[1]{{\color{blue}#1}}
\newcommand{\ot}[1]{{\color{black}#1}}
\newcommand{\ns}[1]{{\color{red!50!gray!60!white}#1}}
\begin{equation*}
\label{eq:row-construction}
\def\arraystretch{1.1}
\begin{blockarray}{cccccccccccccccc}
&& \BAmulticolumn{6}{c}{\text{\small items}} &&&&&&& \\[0.5ex]
r^* & \begin{varwidth}{4em}\begin{center}\small true ratings\end{center}\end{varwidth} & \ns{1} & \ob{0.8} & \ns{0.6} & \ns{0.4} & \ob{0.2} & \ns{0.1} 
&&
T^* &\ot{1}&\ot{1}&\ot{1}&\ot{0}&\ot{0}&\ot{0} \\
\begin{block}{cc[cccccc]cc[cccccc]}
\multirow{6}{*}{$\A$} &
 \multirow{2}{*}{\begin{varwidth}{4em}\begin{center}\small good raters\end{center}\end{varwidth}} & \ob{.9} & \ns{.8} & \ns{.7} & \ns{.6} & \ob{.5} & \ns{.4} \topstrut 
& \multirow{6}{*}{$\implies$} &
\multirow{6}{*}{$M^*$} 
 & \ot{1} & \ot{1} & \ot{1} & \ot{0} & \ot{0} & \ot{0} \topstrut \\
 & & \ns{1} & \ob{.9} & \ns{.8} & \ns{.2} & \ob{.1} & \ns{0} 
 &&
 & \ot{1} & \ot{1} & \ot{1} & \ot{0} & \ot{0} & \ot{0} \\
\cmidrule(lr){3-8}
& \begin{varwidth}{4em}\begin{center}\small random\end{center}\end{varwidth}
& \ob{1} & \ns{0} & \ns{1} & \ns{0} & \ns{1} & \ob{0} 
&&
 & \ot{1} & \ot{0} & \ot{1} & \ot{0} & \ot{1} & \ot{0} \\
\cmidrule(lr){3-8}
& \multirow{2}{*}{\begin{varwidth}{4em}\begin{center}\small adversaries\end{center}\end{varwidth}}
 & \ns{1} & \ns{.8} & \ob{.6} & \ob{0} & \ns{0} & \ns{1} 
&&
 & \ot{1} & \ot{1} & \ot{0} & \ot{0} & \ot{0} & \ot{1} \\
&& \ns{1} & \ns{.8} & \ns{.6} & \ob{0} & \ns{0} & \ob{1} \botstrut 
&&
 & \ot{1} & \ot{1} & \ot{0} & \ot{0} & \ot{0} & \ot{1} \botstrut \\
\end{block}
\end{blockarray}
\end{equation*}
\vskip -0.15in
\caption{Illustration of our problem setting. We observe a small number of 
ratings from each rater (indicated a blue), which we represent as entries in a 
matrix $\A$ (unobserved ratings in red, treated as zero by our algorithm). 
We also rate a small number of items ourself, indicated by $\robs$. Our goal is 
to recover the set $T^*$ representing the top $\beta$ fraction of items under 
our rating. As an intermediate step, we recover a matrix $M^*$ that approximates 
the top items for each individual rater.
}
\label{fig:matrix}
\end{figure}

\begin{theorem}
\label{thm:intro}
In the setting above, suppose 
$k \geq \Omega(1/\beta\alpha^3\epsilon^4)$ and 
$k_0 \geq \Omega(\log(1/\alpha\beta\epsilon)/\beta\epsilon^2)$. Then, with probability 
at least $99\%$, we can identify $\beta n$ items with average quality at most 
$\epsilon$ worse than $T^*$.
\end{theorem}
Amazingly, the amount of work that each worker (and we ourselves) has 
to do does not grow with $n$; it depends only on the fraction $\alpha$ of 
reliable workers and the the desired accuracy $\epsilon$.   While the number 
of evaluations $k$ for each worker is likely not optimal, we note that 
the amount of work $k_0$ required of us is close to optimal: 
for $\alpha \le \beta$, it is information theoretically necessary for us to evaluate $\Omega(1/\beta\eps^2)$ items,via a reduction to estimating 
noisy coin flips \citepa{mannor2004sample}.

Why is it necessary to include some of our own ratings? 
If we did not, and $\alpha < \frac{1}{2}$, then an adversary could create a set of 
dishonest raters that were identical to the reliable raters except with the 
item indices permuted by a random permutation of $\{1,\ldots,m\}$. In this case, 
there is no way to distinguish the honest from the dishonest raters except by 
breaking the symmetry with our own ratings.

Our main result holds in a considerably more general setting where we require a weaker form of inter-rater agreement --- for example, our results hold even if some of the reliable raters are harsher than others, as long as 
the expected ratings induce approximately the same ranking.
The focus on quantiles rather than raw ratings is what enables this. 
Note that once we estimate the quantiles, we can approximately recover the 
ratings by evaluating a few items in each quantile.

Our technical tools draw on semidefinite programming methods for matrix 
completion, which have been used to study graph clustering as well 
as community detection in the stochastic block model \citepm{holland1983stochastic,condon2001algorithms}. 
Our setting corresponds to the sparse case where all nodes have constant degree, 
which has recently seen great interest \citepa{decelle2011asymptotic,
mossel2012stochastic,mossel2013proof,mossel2013belief,
massoulie2014community,guedon2014community,mossel2015consistency,
chin2015stochastic,abbe2015community,makarychev2015learning}. 
\citetm{makarychev2015learning} in particular provide an algorithm that is 
robust to adversarial perturbations, but only if the perturbation has 
size $o(n)$; see also \citetm{cai2015robust} for robusness results when 
the node degree is logarithmic.

Several authors have considered semirandom settings for graph clustering, which 
allow for some types of adversarial behavior \citepa{feige2000finding,
feige2001heuristics,coja2004coloring,krivelevich2006semirandom,
coja2007solving,makarychev2012approximation,chen2014improved,guedon2014community,
moitra2015robust,agarwal2015multisection}. 
In our setting, these semirandom models would need to assume that the adversaries 
are strictly dominated by the reliable raters, in the sense of having lower 
expected accuracy on every item; this is implausible as it rules out 
most types of strategic behavior.
In removing this assumption, we face a key technical challenge: while previous 
analyses consider errors relative to a ground truth clustering, 
in our setting 
the ground truth only exists for rows of the matrix corresponding to reliable 
raters while the remaining rows could behave arbitrarily even in the limit 
where all ratings are observed. This necessitates a more careful analysis, 
which helps to clarify what properties of a clustering are truly necessary 
for identifying it.


\section{Algorithm and Intuition}
\label{sec:algorithm}

\begin{algorithm}[b!]
\caption{Algorithm for recovering $\beta$-quantile matrix $\M$ using
(unreliable) ratings $\A$.}
\label{alg:recover-M}
\begin{algorithmic}[1]
\State Parameters: reliable fraction $\alpha$, quantile $\beta$, tolerance $\epsilon$, number of raters $n$, number of items $m$
\State Input: noisy rating matrix $\A$
\State Let $\M$ be the solution of the optimization problem \eqref{eq:optimization-noisy}:
  \begin{align}
  \label{eq:optimization-noisy}
  \text{maximize } &\langle \A, M \rangle, \\
  \notag \text{ subject to } &0 \leq M_{ij} \leq 1 \,\,\, \forall i,j, \\
  \notag                     &{\textstyle \sum_{j}} M_{ij} \leq \beta m \,\,\, \forall j, \quad\quad
                      \|M\|_* \leq \frac{2}{\alpha\epsilon}\sqrt{\alpha\beta nm}, \phantom{xxxxxxx}
  \end{align}
  where $\|\cdot\|_*$ denotes nuclear norm.
\State Output $\M$.
\end{algorithmic}
\end{algorithm}

We now describe our recovery algorithm. To fix notation, we assume that 
there are $n$ raters and $m$ items, and that we observe a matrix 
$\A \in [0,1]^{n \times m}$: 
$\A_{ij} = 0$ if rater $i$ does not rate item $j$, and otherwise $\A_{ij}$ 
is the assigned rating, which takes values 
in $[0,1]$. In the settings we 
care about $\A$ is very sparse --- each rater only rates a few items.
Remember that our goal is to recover the $\beta$-quantile $T^*$ of the 
best items according to our own rating. 

Our algorithm is based on the following intuition: the reliable raters must 
(approximately) agree on the ranking of items, and so if we can cluster the 
rows of $\A$ appropriately, then the reliable raters should form a single very large cluster 
(of size $\alpha n$). There can be at most $\frac{1}{\alpha}$ disjoint clusters of this size, and 
so we can manually check the accuracy of each large cluster (by checking agreement 
with our own rating on a few randomly selected items) and 
then choose the best one.

One major challenge in using the clustering intuition is the sparsity of 
$\A$: any two rows of $\A$ will almost certainly have no ratings in common, 
so we must exploit the global structure of $\A$ to discover 
clusters, rather than using pairwise comparisons of rows.
The key is to view our problem as a form of \emph{noisy matrix completion} --- 
we imagine a matrix $\Aavg$ in which all the ratings have been filled in 
and all noise from individual ratings has been removed. We define a 
matrix $\Mavg$ that indicates the top $\beta m$ items in each row of $\Aavg$: 
$\Mavg_{ij} = 1$ if item $j$ has one of the top $\beta m$ ratings from rater $i$, 
and $\Mavg_{ij} = 0$ otherwise (this differs from the actual 
definition of $\Mavg$ given in Section~\ref{sec:approach-M}, but is the same in spirit). If we could recover $\Mavg$, we would be 
close to obtaining the clustering we wanted.

The key observation that allows us to approximate $\Mavg$ given only the noisy, 
incomplete $\A$ is that \emph{$\Mavg$ has low-rank structure}: since all 
of the reliable raters agree with each other, their rows in $\Mavg$ are all 
identical, and so there is an $(\alpha n) \times m$ submatrix of $\Mavg$ with 
rank $1$. This inspires the low-rank matrix completion algorithm for recovering 
$\M$ given in Algorithm~\ref{alg:recover-M}. Each row of $M$ is constrained 
to have sum at most $\beta m$, and $M$ as a whole is constrained to have 
nuclear norm $\|M\|_*$ at most $\frac{2}{\alpha \epsilon}\sqrt{\alpha\beta nm}$. 
Recall that the \emph{nuclear norm} is the sum of the singular values of 
$M$; in the same way that the $\ell^1$-norm is a convex surrogate for the 
$\ell^0$-norm, the nuclear norm acts as a convex surrogate for the rank of $M$ 
(i.e., number of non-zero singular values). The optimization problem 
\eqref{eq:optimization-noisy} therefore chooses a set of $\beta m$ items in each 
row to maximize the corresponding values in $\A$, while constraining the item 
sets to have low rank (where low rank is relaxed to low nuclear norm to obtain 
a convex problem). 
This low-rank constraint acts as a strong regularizer that quenches the noise 
in $\A$.

\algdef{SE}[DOWHILE]{Do}{doWhile}{\algorithmicdo}[1]{\algorithmicwhile\ #1}%
\begin{algorithm}[t!]
\caption{Algorithm for recovering an accurate $\beta$-quantile $T$ from the $\beta$-quantile matrix $\M$.}
\label{alg:recover-T}
\begin{algorithmic}[1]
\State Parameters: tolerance $\epsilon$, reliable fraction $\alpha$
\State Input: matrix $\M$ of approximate $\beta$-quantiles, noisy ratings $\robs$, $\robs'$
\State Let $\goodapprox$ be the set of $\alpha n$ indices $i \in [n]$ for which 
       $\sum_j \M_{ij}\robs_j$ is largest.
\State $T_0 \gets \frac{1}{|\goodapprox|} \sum_{i \in \goodapprox} \M_i$. \Comment{$T_0 \in [0,1]^m$}
\State {\bfseries do} \ $T \gets \textsc{RandomizedRound}(T_0)$ {\bfseries while } $\langle T-T_0, \robs' \rangle < -\frac{\epsilon}{4}\beta k$
\State \Return $T$ \Comment{$T \in \{0,1\}^m$} 
\end{algorithmic}
\end{algorithm}

Once we have recovered $\M$ using Algorithm~\ref{alg:recover-M}, it remains to 
recover a specific set $T$ that approximates the $\beta$-quantile according to 
our ratings. Algorithm~\ref{alg:recover-T} provides a recipe for doing so: 
first, rate $k_0$ items at random, obtaining the vector $\robs$:
$\robs_j = 0$ if we did not rate item $j$, and otherwise $\robs_j$ is 
the (possibly noisy) rating that we assign to item $j$. Next, score each 
row $\M_{i}$ based on the noisy ratings $\sum_j \M_{ij}\robs_j$, and let 
$T_0$ be the average of the $\alpha n$ highest-scoring $\M_i$.
Finally, use randomized rounding to turn the vector $T_0 \in [0,1]^m$ into 
a discrete vector $T \in \{0,1\}^m$, and treat $T$ as the indicator function 
of a set approximating the $\beta$-quantile
(see Section~\ref{sec:rounding} for details of the rounding algorithm).

In summary, given a noisy rating matrix $\Aobs$, we will first run 
Algorithm~\ref{alg:recover-M} to recover a $\beta$-quantile matrix $\M$ for 
each rater, and then run Algorithm~\ref{alg:recover-T} to recover our 
personal $\beta$-quantile from $\M$.

\paragraph{Possible attacks by adversaries.} In our algorithm, 
the adversaries can influence $\M_i$ for reliable raters $i$ via 
the nuclear norm constraint (note that the other constraints are 
separable across rows). This makes sense because 
the nuclear norm is what causes us to pool global structure across 
raters (and thus potentially pool bad information). In order to 
limit this influence, the constraint on the 
nuclear norm is weaker than is typical by a factor of $\frac{2}{\epsilon}$; 
it is not clear to us whether this is actually necessary or due to a 
loose analysis. (Note that $M_{\good}^*$--$M^*$ restricted to the 
reliable rows--has nuclear norm $\sqrt{\alpha\beta nm}$, since it is the 
$\alpha n \times \beta m$ all-$1$s matrix padded by zeros; the constraint on $\|M\|_*$ must 
be at least $\frac{1}{\alpha}$ times as large as this since the adversaries 
could produce $\frac{1}{\alpha}$ permuted copies of $M_{\good}^*$.)

The constraint $\sum_j M_{ij} \leq \beta m$ is 
not typical in the literature. For instance, \citepm{chen2014improved} 
place no 
constraint on the sum of each row in $M$ (instead of recovering 
the $\beta$-quantile, they normalize $\A$ to lie in $[-1,1]^{m \times m}$ and recover 
the items with a positive rating).
Our row normalization constraint prevents an attack 
in which a spammer rates a random subset of items as high as possible and 
rates the remaining items as low as possible. If the actual set of 
high-quality items has density much smaller than $50\%$, then the 
spammer gains undue influence relative to honest raters that 
only rate e.g. $10\%$ of the items highly. Normalizing $M$ to 
have a fixed row sum prevents this; see Section~\ref{sec:adversary-examples} 
for details.

\section{Assumptions and Approach}
\vskip -0.08in
\label{sec:assumptions}
\label{sec:approach}

We now state our assumptions more formally, state the general form 
of our results, and outline the key ingredients of the proof.
In our setting, we can query a rater $i \in [m]$ and item $j \in [m]$ to 
obtain a rating $\A_{ij} \in [0,1]$. Let 
$\ravg \in [0,1]^m$ denote the vector of true ratings of the items. 
We can also query an item $j$ (by rating it ourself) to obtain a noisy 
rating $\robs_j$ such that $\bE[\robs_j] = \ravg_j$.

Let $\good \subseteq [n]$ be the set of reliable raters, where $|\good| \geq \alpha n$.
Our main assumption is that the reliable raters make independent errors:
\begin{assumption}[Independence]
\label{ass:independent}
When we query a pair $(i,j)$, and $i \in \good$, we obtain an output 
$\A_{ij}$ whose value is independent of all of the other queries so far.
Similarly, when we query an item $j$, we obtain an output $\robs_j$ that 
is independent of all of the other queries so far.
\end{assumption}
Note that Assumption~\ref{ass:independent} allows the unreliable ratings to 
depend on all previous ratings and also allows arbitrary collusion 
among the unreliable raters. 
In our algorithm, we will generate our own ratings after querying everyone 
else, which ensures that at least $\robs$ is independent of the adversaries.

We need a way to formalize the idea that the reliable raters 
agree with us. To this end, for $i \in \good$ let $\Aavg_{ij}$ be
the expected rating that rater $i$ assigns to item $j$.
We want $\Aavg$ to be roughly increasing in $\ravg$:
\begin{definition}[Monotonic raters]
\label{def:lipschitz}
We say that the reliable raters are \emph{$(L,\epsilon)$-monotonic} if 
\vskip -0.18in
\begin{equation}
\label{eq:lipschitz}
\ravg_j - \ravg_{j'} \leq L \cdot (\Aavg_{ij} - \Aavg_{ij'}) + \epsilon
\end{equation}
\vskip -0.07in
whenever $\ravg_j \geq \ravg_{j'}$, and
for all $i \in \good$ and all $j,j' \in [m]$.
\end{definition}
The $(L,\epsilon)$-monotonicity property says that if we think that one item is 
substantially better than another item, the reliable raters should think 
so as well. As an example, suppose that our own ratings are binary 
($\ravg_j \in \{0,1\}$) and that each rating $\Aobs_{i,j}$ matches $\ravg_j$ 
with probability $\frac{3}{5}$. Then 
$\Aavg_{i,j} = \frac{2}{5} + \frac{1}{5}\ravg_j$, 
and hence the ratings are $(5,0)$-monotonic. 
In general, the monotonicity property is fairly mild --- if the reliable ratings 
are not $(L,\epsilon)$-monotonic, it is not clear that they should 
even be called reliable!

\begin{algorithm}[t!]
\caption{Algorithm for obtaining (unreliable) ratings matrix $\A$ and noisy 
ratings $\robs$, $\robs'$.}
\label{alg:create-A}
\begin{algorithmic}[1]
\State Input: number of raters $n$,  number of items $m$, and ratings per rater $k$.
\State Initially assign each rater to each item independently with probability $k/m$.  
\State For every rater assigned more than $2k$ items, un-assign items until 
       there are $2k$ remaining.
\State For every item assigned to more than $2k$
       raters, un-assign raters until there are $2k$ remaining.
\State Have the raters submit ratings of their assigned items, and let $\A$ 
       denote the resulting matrix of ratings with missing entries fill in with 
       zeros.
\State Generate each of $\robs$ and $\robs'$ by rating items with probability $\frac{k_0}{m}$ (fill in missing entries with zeros)
\State Output $\A$, $\robs$, and $\robs'$
\end{algorithmic}
\end{algorithm}

\paragraph{Algorithm for collecting ratings.}
Under the model given in Assumption~\ref{ass:independent}, 
our algorithm for collecting ratings is given in 
Algorithm~\ref{alg:create-A}. Given integers $k$ and $k_0$, 
Algorithm~\ref{alg:create-A} assigns each rater at most $2k$ 
ratings, and assigns us $2k_0$ ratings in expectation. The output is a 
noisy rating matrix $\A \in [0,1]^{n \times m}$ as well 
as noisy rating vectors $\robs, \robs' \in [0,1]^m$ (we need 
to create two independent rating vectors for technical reasons; 
in practice we can use a single vector).
Our main result states that we can use $\A$ and $\robs$ to 
estimate the $\beta$-quantile $T^*$; throughout we will assume 
that $m$ is at least $n$.
\begin{theorem}
\label{thm:main}
Let $m \geq n$. 
Suppose that Assumption~\ref{ass:independent} holds, that 
the reliable raters are $(L,\epsilon_0)$-monotonic, and 
that we run Algorithm~\ref{alg:create-A} to obtain noisy ratings, 
with $k \geq \Omega\p{\frac{\log^3(1/\delta)}{\beta\alpha^3\epsilon^4}\frac{m}{n}}$ and
$k_0 \geq \Omega\p{\frac{\log(1/\alpha\beta\epsilon\delta)}{\beta\epsilon^2}}$.
Then, with probability $1-\delta$, 
Algorithms~\ref{alg:recover-M} and \ref{alg:recover-T} recover a set $T$ satisfying 
\vskip -0.18in
\[ \frac{1}{\beta m} \p{\sum_{j \in T^*} \ravg_j - \sum_{j \in T} \ravg_j} 
\leq (L+1) \cdot \epsilon + \epsilon_0. \]
\vskip -0.07in
\end{theorem}
Note that the amount of work for the raters scales as $\frac{m}{n}$. Some dependence 
on $\frac{m}{n}$ is necessary, since we need to make sure that every item gets rated at least once.

The proof of Theorem~\ref{thm:main} can be split into two parts: analyzing 
Algorithm~\ref{alg:recover-M} (Section~\ref{sec:approach-M}), 
and analyzing Algorithm~\ref{alg:recover-T} (Section~\ref{sec:approach-T}). 
At a high level, analyzing Algorithm~\ref{alg:recover-M} involves showing that 
the nuclear norm constraint in \eqref{eq:optimization-noisy} imparts sufficient 
noise robustness while not allowing the adversary too much influence over the 
reliable rows of $\M$. Analyzing Algorithm~\ref{alg:recover-T} is far more 
straightforward, 
and requires only standard concentration inequalities and a standard randomized 
rounding idea (though the latter is perhaps not well-known, so we will explain 
it briefly in Section~\ref{sec:approach-T}).

\section{Recovering $\M$ (Algorithm~\ref{alg:recover-M})}
\label{sec:approach-M}

The goal of this section is to show that solving the optimization 
problem \eqref{eq:optimization-noisy} recovers a matrix $\M$ that 
approximates the $\beta$-quantile of $\ravg$ in the following sense:
\begin{proposition}
\label{prop:recover-M}
Under the conditions of Theorem~\ref{thm:main}, Algorithm~\ref{alg:recover-M} 
outputs a matrix $\M$ satisfying 
\[ \frac{1}{|\good|} \frac{1}{\beta m} \sum_{i \in \good} \sum_{j \in [m]} (T^*_j - \M_{i,j})\Aavg_{ij} \leq \epsilon, \]
where $T^*_j = 1$ if $j$ lies in the $\beta$-quantile of $\ravg$, and is $0$ otherwise.
\end{proposition}
Proposition~\ref{prop:recover-M} says that the row $\M_i$ 
is good according to rater $i$'s ratings $\Aavg_i$. Note that
$(L,\epsilon_0)$-monotonicity 
then implies that $\M_i$ is also good according to $\ravg$.
In particular (see \ref{sec:lipschitz-details} for details)
\begin{align}
\label{eq:lipschitz-usage}
\frac{1}{|\good|}\frac{1}{\beta m}\sum_{i \in \good}\sum_{j \in [m]} (T_j^*-\M_{ij})\ravg_j
&\leq L \cdot \frac{1}{|\good|}\frac{1}{\beta m}\sum_{i \in \good}\sum_{j \in [m]} (T_j^*-\M_{ij})\Aavg_{ij} + \epsilon_0 
\leq L \cdot \epsilon + \epsilon_0.
\end{align}

Proving Proposition~\ref{prop:recover-M} involves two major steps: showing 
(a) that the nuclear norm constraint in \eqref{eq:optimization-noisy} 
imparts noise-robustness, and (b) that the constraint does not allow 
the adversaries to influence $\M_{\good}$ too much. (For a matrix $X$ 
we let $X_{\good}$ denote the rows indexed by $\sC$ and $X_{\bad}$ the remaining rows.)

In a bit 
more detail, if we let $\Mm$ denote a denoised version of $\M$, and $\Aa$ 
denote a denoised version of $\A$, we first show 
(Lemma~\ref{lem:objective-bound}) that 
$\langle B, \M - \Mm \rangle \geq -\epsilon'$ for some $\epsilon'$ 
determined below. This is established via the matrix concentration 
inequalities in \citetm{le2015concentration}. Lemma~\ref{lem:objective-bound} already 
suffices for standard approaches \citepm[e.g.,][]{guedon2014community}, 
but in our case we must grapple with the issue that the rows of $B$ could be 
arbitrary outside of $\sC$, and hence closeness according to $B$ may not 
imply actual closeness between $\M$ and $\Mm$. Our main 
technical contribution, Lemma~\ref{lem:subgradient}, shows
that $\langle B_{\good}, \M_{\good} - \Mm_{\good} \rangle \geq \langle B, \M - \Mm \rangle - \epsilon'$; 
that is, \emph{closeness according to $B$ implies closeness according to 
$B_{\good}$}. We can then restrict attention to the 
reliable raters, and obtain Proposition~\ref{prop:recover-M}.

\paragraph{Part 1: noise-robustness.} Let $\Aa$ be the matrix satisfying 
$\Aa_{\good} = \frac{k}{m}\Aavg_{\good}$, $\Aa_{\bad} = \Aobs_{\bad}$, 
which denoises $\A$ on $\good$.
The scaling $\frac{k}{m}$ is chosen so that 
$\bE[\Aobs_{\good}] \approx \Aa_{\good}$.
Also define $R \in \bR^{n \times m}$ by $R_{ij} = T_j^*$.

Ideally, we would like to have $M_{\good} = R_{\good}$, i.e., $M$ matches $T^*$ on 
all the rows of $\good$. In light of this, 
we will let $\Mm$ be the solution to the following ``corrected'' program, which 
we don't have access to (since it involves knowledge of $\Aavg$ and $\good$), 
but which will be useful for analysis purposes:
\begin{align}
\label{eq:optimization-noiseless}
\text{maximize } &\langle \Aa, M \rangle, \\
\notag \text{ subject to } &M_{\good} = R_{\good}, 
  &&\hskip-0.4in 0 \leq M_{ij} \leq 1 \,\,\, \forall i,j,  \\
\notag  &{\textstyle \sum_j} M_{ij} \leq \beta m \,\,\, \forall i, 
  &&\hskip-0.4in \|M\|_* \leq \frac{2}{\alpha\epsilon}\sqrt{\alpha\beta nm} \phantom{xxxxxxx}
\end{align}
Importantly, \eqref{eq:optimization-noiseless} enforces $\Mm_{ij} = T_j^*$ for all 
$i \in \good$. Lemma~\ref{lem:objective-bound} shows that $\M$ is ``close'' to $\Mm$: 
\begin{lemma}
\label{lem:objective-bound}
Let $m \geq n$. Suppose that Assumption~\ref{ass:independent} holds and that 
$k = \Omega\p{ \frac{\log^3(1/\delta)}{\beta\alpha^3\eps^4}\frac{m}{n}}$. 
Then, the solution $\M$ to \eqref{eq:optimization-noisy} performs nearly as 
well as $\Mm$ under $B$; specifically, with probability $1-\delta$,
\begin{equation}
\label{eq:objective-bound}
\langle \Aa, \M \rangle \geq \langle \Aa, \Mm \rangle - \epsilon \alpha\beta kn.
\end{equation}
\end{lemma}
Note that $\M$ is not necessarily feasible for \eqref{eq:optimization-noiseless}, 
because of the constraint $M_{\good} = R_{\good}$; Lemma~\ref{lem:objective-bound} 
merely asserts that $\M$ approximates $\Mm$ in objective value. The proof of 
Lemma~\ref{lem:objective-bound}, given in Section~\ref{sec:objective-bound-proof}, 
primarily involves establishing a 
\emph{uniform deviation result}; if we let $\sP$ denote the feasible set for 
\eqref{eq:optimization-noisy}, then we wish to show that 
$|\langle \A - B, M \rangle| \leq \frac{1}{2}\epsilon \alpha\beta kn$ for all 
$M \in \sP$. This would imply that the objectives of 
\eqref{eq:optimization-noisy} and \eqref{eq:optimization-noiseless} are 
essentially identical, and so optimizing one also optimizes the other.

Using the inequality $|\langle \A - B, M \rangle| \leq \|\A-B\|_{\op}\|M\|_*$, 
where $\|\cdot\|_{\op}$ denotes operator norm, it suffices to establish a matrix 
concentration inequality bounding $\|\A - B\|_{\op}$.   
This bound follows from the general matrix concentration result of~\cite{le2015concentration},
stated in Section~\ref{sec:le-statement}.

\paragraph{Part 2: bounding the influence of adversaries.} 
We next show that the nuclear norm constraint does 
not give the adversaries too much influence over the de-noised program 
\eqref{eq:optimization-noiseless}; this is the most novel aspect 
of our argument.

\begin{figure}
\begin{center}
\begin{tikzpicture}[scale=5/6, every node/.style={scale=5/6}]
\def\s{6.7}
\def\x{0.3*\s};
\def\y{0.2*\s};
\def\X{1*\s};
\def\Y{0.9*\s};
\def\pux{0.57*\s};
\def\puy{\Y};
\def\Msx{\X};
\def\Msy{0.7*\s};
\def\Bx{\Msx+0.3*\s};
\def\By{\Msy+0.12*\s};
\def\plx{0.8*\s};
\def\ply{\y};
\def\ex{\X-0.14*\s}
\coordinate (O) at (\x,\y);
\coordinate (Cx) at (\X,\y);
\coordinate (Cy) at (\x,\Y);
\coordinate (C) at (\X,\Y);
\coordinate (x-ax) at (\X+0.4,\y);
\coordinate (y-ax) at (\x,\Y+0.4);
\coordinate (p-l) at (\plx,\ply);
\coordinate (p-u) at (\pux,\puy);
\coordinate (M-star) at (\Msx,\Msy);
\coordinate (B) at (\Bx,\By);
\def\Bpx{{\Bx-(\By-\Msy)/(\Msx-\pux)*(\puy-\Msy)}};
\def\Bpy{\Msy};
\coordinate (Bp) at (\Bpx,\Bpy);
\draw[opacity=0,name path=l--c] (M-star) -- (p-l);
\draw[opacity=0,name path=l--b] (p-u) -- (M-star);
\draw[opacity=0,name path=l--a] (\ex,\X) -- (\ex,\x);
\path [name intersections={of=l--a and l--b,by=e1}];
\path [name intersections={of=l--a and l--c,by=e2}];
\fill [opacity=0.3,gray] (O) -- (Cy) -- (C) -- (Cx) -- cycle;
\fill [opacity=0.3,blue]
  (O) -- (Cy) -- (p-u) -- (M-star) -- (p-l) -- cycle;
\draw (p-u) edge[thick,blue!70!black] node[pos=0.3,below,sloped] {$\|M\|_* \leq \rho$} (M-star);
\fill [opacity=0.3,red] (e1) -- (M-star) -- (e2) -- cycle;
\draw (M-star) edge[thick,blue!70!black] (p-l);
\draw (O) edge[thick,->] (x-ax);
\draw (O) edge[thick,->] (y-ax);
\node[below=1.2em of Cx] (Cx-s) {};
\node[above=1.2em of C] (C-n) {$M_{\good} = R_{\good}$};
\draw (C-n) edge[dashed] node[pos=0.1,sloped,above] {} (Cx-s);
\draw (Cx) edge (C);
\draw (Cy) edge (C);
\fill(M-star) circle(1.5pt);
\draw (M-star) edge[->] (B);
\draw (M-star) edge[->] (Bp);
\node[right=-0.1em of B] {$B$};
\node[right=-0.1em of Bp] {$B_{\good}\!-\!Z_{\good}$};
\node[right=-0.1em of x-ax] {$M_{\good}$};
\node[above=-0.1em of y-ax] {$M_{\bad}$};
\node[below left=-0.3em and -0.1em of M-star,scale=0.9] {$M^*$};
\draw (\ex,\Y+0.08*\s) edge[dashed,red!70!black] node[pos=0.47,above,rotate=90,scale=0.68,red!50!black] {$\langle B_{\good}\!-\!Z_{\good}, M^*\!-\!M \rangle \leq \epsilon$} (\ex,\y-0.09*\s);
\coordinate[below right=0.65em and 0.45em of e1] (Mt);
\fill(Mt) circle(1.5pt);
\node[above right=0em and 0em of Mt,scale=0.9] {$\M$};
\end{tikzpicture}
\end{center}
\vskip -0.15in
\caption{Illustration of our Lagrangian duality argument, and of the role 
of $Z$. The blue region represents the nuclear norm constraint and the gray 
region the remaining constraints. Because the blue region slopes downwards, 
a decrease in $M_{\good}$ can be offset by an increase in $M_{\bad}$ when 
measuring $\langle B, M \rangle$. The vector $B-Z$ accounts for this offset, 
and the red region represents the constraint 
$\langle B_{\good}-Z_{\good}, M^*_{\good} - M_{\good} \rangle \leq \epsilon$, which is guaranteed to contain $\M$.
}
\vskip 0.05in
\label{fig:lagrangian}
\end{figure}
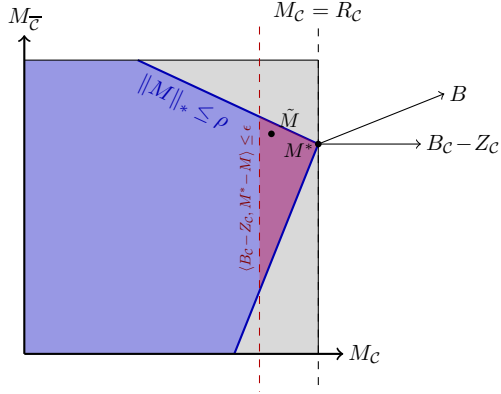

Suppose that the constraint on $\|M\|_*$ were not present in 
\eqref{eq:optimization-noiseless}. Then the adversaries would have 
no influence on $\Mm_{\good}$, because all the remaining constraints 
in \eqref{eq:optimization-noiseless} are separable across rows. 
How can we quantify the effect of this nuclear norm constraint?
We exploit Lagrangian duality, which allows us to replace constraints 
with appropriate modifications to the objective function.


To gain some intuition, consider 
Figure~\ref{fig:lagrangian}. The key is that the Lagrange 
multiplier $Z_{\good}$ can bound the amount that $\langle B, M \rangle$ 
can increase due to changing $M$ outside of $\good$.
If we formalize this and analyze $Z$ in detail, we obtain the 
following result:
\begin{lemma}
\label{lem:subgradient}
Let $m \geq n$. Suppose that $k = \Omega\p{\frac{\log^3(1/\delta)}{\alpha\beta\epsilon^2}\frac{m}{n}}$. 
Then with probability at least $1-\delta$ there exists a matrix $Z$ with 
$\rank(Z) = 1$, $\|Z\|_F \leq \epsilon k\sqrt{\alpha\beta n/m}$ such that
\begin{align}
\label{eq:localize}
\langle \Aa_{\good} - Z_{\good}, \Mm_{\good} - M_{\good} \rangle &\leq \langle \Aa, \Mm - M \rangle \text{ for all $M \in \sP$}.
\end{align}
\end{lemma}
By localizing $\langle \Aa, \Mm - M \rangle$ to $\good$ via 
\eqref{eq:localize}, 
Lemma~\ref{lem:subgradient} bounds the effect that the adversaries can have 
on $\M_{\good}$. It is therefore the key 
technical tool powering our results, and is proved in 
Section~\ref{sec:subgradient-proof}. Proposition~\ref{prop:recover-M} 
is proved from Lemmas~\ref{lem:objective-bound} and \ref{lem:subgradient} 
in Section~\ref{sec:recover-M-proof}.

\section{Recovering $T$ (Algorithm~\ref{alg:recover-T})}
\vskip -0.10in
\label{sec:approach-T}
\label{sec:rounding}

In this section we show that if $\M$ satisfies the conclusion of 
Proposition~\ref{prop:recover-M}, then Algorithm~\ref{alg:recover-T} 
recovers a set $T$ that approximates $T^*$ well. Formally, we show 
the following:
\begin{proposition}
\label{prop:recover-T}
Suppose Assumption~\ref{ass:independent} holds and 
$k_0 \geq \Omega\p{\frac{\log(1/\alpha\beta\epsilon\delta)}{\beta\epsilon^2}}$. 
With probability $1-\delta$, Algorithm~\ref{alg:recover-T} outputs a set $T$ satisfying 
\begin{equation}
\label{eq:recover-T}
\frac{1}{\beta m}\sum_{j \in T} \ravg_j \geq \p{\frac{1}{\beta m}\frac{1}{|\good|}\sum_{i \in \good} \sum_{j \in [m]} \M_{ij}\ravg_j} - \epsilon.
\end{equation}
\end{proposition}
The validity of this procedure hinges on two results. First, establish 
a concentration bound showing that $\sum_j \M_{ij}\robs_j$ is close to 
$\frac{k_0}{m}\sum_j \M_{ij}\rtrue_j$ for all $i \in \good$, which implies that 
the $\goodfrac n$ best rows of $\M$ according to $\robs$ also look good 
according to $\ravg$. This yields the following lemma:
\begin{lemma}
\label{lem:robs-rtrue}
Let $\goodapprox$ be the $\goodfrac n$ best rows according to $\robs$, as 
in Algorithm~\ref{alg:recover-T}. 
Suppose that $\robs$ satisfies Assumption~\ref{ass:independent} and that
$k_0 \geq \Omega\p{\frac{\log(1/\alpha\failprob)}{\beta\epsilon^2}}$. 
Then, with probability $1-\delta$, we have
\begin{equation}
\label{eq:robs-rtrue}
\frac{1}{\goodfrac n} \sum_{i \in \goodapprox} \Big(\sum_{j \in [m]} \M_{ij}\rtrue_j\Big) \geq \frac{1}{|\good|} \sum_{i \in \good} \Big(\sum_{j \in [m]} \M_{ij}\rtrue_j\Big) - \frac{\epsilon}{4}\beta m.
\end{equation}
\end{lemma}
See Section~\ref{sec:concentration-proof} for a proof.
The idea is to establish a uniform bound showing that 
$\sum_{i \in S} \sum_{j \in [m]} \M_{ij}(\robs_j - \frac{k_0}{m}\rtrue_j)$ is small for any 
set of $\goodfrac n$ rows $S$, and hence that greedily taking the $\goodfrac n$ 
best rows according to $\robs$ is almost as good as taking the $\goodfrac n$ 
best rows according to $\rtrue$. We 
improve over a na\"{i}ve union bound by exploiting power mean 
inequalities on cumulant functions. 

Having recovered a set $\goodapprox$ of good rows, define their 
average $T_0 \in [0,1]^m$ as $T_0 \eqdef \frac{1}{|\goodapprox|} \sum_{i \in \goodapprox} \M_i$. 
We need to turn $T_0$ into a binary vector so that 
Algorithm~\ref{alg:recover-T} can output a set;
we do so via randomized rounding, obtaining a vector $T \in \{0,1\}^m$ such that 
$\bE[T_0] = T$.
Our rounding procedure is given in Algorithm~\ref{alg:round}; the following 
lemma, proved in \ref{sec:rounding-proof}, asserts its correctness:
\begin{lemma}
\label{lem:rounding}
The output $T$ of Algorithm~\ref{alg:round} satisfies $\bE[T] = T_0$, 
$\|T\|_0 \leq \beta m$.
\end{lemma}

\begin{algorithm}[h!]
\caption{Randomized rounding algorithm.}
\label{alg:round}
\begin{algorithmic}[1]
\Procedure{RandomizedRound}{$T_0$}
\Comment{$T_0 \in [0,1]^m$ satisfies $\|T_0\|_1 \leq \beta m$}
\State Let $s$ be the vector of partial sums of $T_0$ \Comment i.e., $s_j = (T_0)_1 + \cdots + (T_0)_j$
\State Sample $u \sim \Uniform([0,1])$.
\State $T \gets [0,\ldots,0] \in \bR^m$
\For{$z = 0$ {\bfseries to} $\beta m - 1$}
\State Find $j$ such that $u+z \in [s_{j-1},s_j)$. \Comment{if no such $j$ exists, skip the next step}
\State $T_j \gets 1$
\EndFor
\State \Return $T$
\EndProcedure
\end{algorithmic}
\end{algorithm}

The remainder of the proof involves lower-bounding the probability 
that $T$ is accepted in each stage of the while loop in 
Algorithm~\ref{alg:recover-T}.
We refer the reader to Section~\ref{sec:recover-T-proof} for details.
%

\section{Open Directions and Related Work}
\label{sec:discussion}

\paragraph{Future Directions}
On the theoretical side, perhaps the most immediate open question is whether it is 
possible to improve the dependence of $k$ (the amount of work required per worker) 
on the parameters $\alpha, \beta,$ and $\eps$.  It is tempting to hope that 
when $m = n$ a tight result would have 
$k = \Theta\p{\frac{\log(1/\alpha)}{\min (\alpha,\beta)\eps^2}}$, in loose analogy
to recent results for the stochastic block model \citepm{banks2016information}.

Our results also leave some open questions for variations on our setting. 
One concerns the regime where $m \ll n$: in this case, can we get by 
with much less work per rater?
Another question concerns adaptivity: if the choice of queries is based 
on previous worker ratings, can we reduce the amount of work?
We would be quite interested in answers to either question.

\paragraph{Related work.}
Our setting is closely related to the problem of \emph{peer prediction} 
\citepm{miller2005eliciting}, in which we wish to obtain truthful information 
from a population of raters by exploiting inter-rater agreement. 
While several mechanisms have been proposed for these tasks, 
they typically assume that rater accuracy is observable online
\citepm{resnick2007influence}, that raters are 
rational agents maximizing a payoff function \citepm{dasgupta2013crowdsourced,
kamble2015truth,shnayder2016strong}, that the workers follow a simple 
statistical model \citepm{karger2014budget,zhang2014crowdsourcing,
zhou2015regularized}, or some combination of the above \citepm{shah2015double,
shah2015approval}. 


The work most close to ours is \citetm{christiano2014provably,
christiano2016robust}, which studies online collaborative prediction in 
the presence of adversaries; roughly, when raters interact with an item 
they predict its quality and afterwards observe the actual quality; the 
goal is to minimize the number of incorrect 
predictions among the honest raters. This differs from our setting in that 
(i) the raters are trying to learn the item qualities as part of the task, 
and (ii) there is no requirement to induce a final global estimate of the 
high-quality items, which is necessary for estimating quantiles.
It seems possible however that there are theoretical ties between this 
setting and ours, which would be interesting to explore.

\nocite{decelle2011asymptotic,
mossel2012stochastic,massoulie2014community,
guedon2014community,abbe2015community}
\nocite{feige2001heuristics,coja2004coloring}
{\small
\setlength{\bibsep}{2.5pt plus 3.0ex}
\bibliographystyle{plainnat}
\bibliography{refdb/all}

\begin{thebibliography}{44}
\providecommand{\natexlab}[1]{#1}
\providecommand{\url}[1]{\texttt{#1}}
\expandafter\ifx\csname urlstyle\endcsname\relax
  \providecommand{\doi}[1]{doi: #1}\else
  \providecommand{\doi}{doi: \begingroup \urlstyle{rm}\Url}\fi

\bibitem[Abbe and Sandon(2015)]{abbe2015community}
E.~Abbe and C.~Sandon.
\newblock Community detection in general stochastic block models: fundamental
  limits and efficient recovery algorithms.
\newblock \emph{arXiv}, 2015.

\bibitem[Agarwal et~al.(2015)Agarwal, Bandeira, Koiliaris, and
  Kolla]{agarwal2015multisection}
N.~Agarwal, A.~S. Bandeira, K.~Koiliaris, and A.~Kolla.
\newblock Multisection in the stochastic block model using semidefinite
  programming.
\newblock \emph{arXiv}, 2015.

\bibitem[Banks and Moore(2016)]{banks2016information}
J.~Banks and C.~Moore.
\newblock Information-theoretic thresholds for community detection in sparse
  networks.
\newblock \emph{arXiv}, 2016.

\bibitem[Cai and Li(2015)]{cai2015robust}
T.~T. Cai and X.~Li.
\newblock Robust and computationally feasible community detection in the
  presence of arbitrary outlier nodes.
\newblock \emph{The Annals of Statistics}, 43\penalty0 (3):\penalty0
  1027--1059, 2015.

\bibitem[Chen et~al.(2014)Chen, Sanghavi, and Xu]{chen2014improved}
Y.~Chen, S.~Sanghavi, and H.~Xu.
\newblock Improved graph clustering.
\newblock \emph{IEEE Transactions on Information Theory}, 60\penalty0
  (10):\penalty0 6440--6455, 2014.

\bibitem[Chin et~al.(2015)Chin, Rao, and Vu]{chin2015stochastic}
P.~Chin, A.~Rao, and V.~Vu.
\newblock Stochastic block model and community detection in the sparse graphs:
  A spectral algorithm with optimal rate of recovery.
\newblock In \emph{Conference on Learning Theory (COLT)}, 2015.

\bibitem[Christiano(2014)]{christiano2014provably}
P.~Christiano.
\newblock Provably manipulation-resistant reputation systems.
\newblock \emph{arXiv}, 2014.

\bibitem[Christiano(2016)]{christiano2016robust}
P.~Christiano.
\newblock Robust collaborative online learning.
\newblock \emph{arXiv}, 2016.

\bibitem[Coja-Oghlan(2004)]{coja2004coloring}
A.~Coja-Oghlan.
\newblock Coloring semirandom graphs optimally.
\newblock \emph{Automata, Languages and Programming}, pages 71--100, 2004.

\bibitem[Coja-Oghlan(2007)]{coja2007solving}
A.~Coja-Oghlan.
\newblock Solving {NP}-hard semirandom graph problems in polynomial expected
  time.
\newblock \emph{Journal of Algorithms}, 62\penalty0 (1):\penalty0 19--46, 2007.

\bibitem[Condon and Karp(2001)]{condon2001algorithms}
A.~Condon and R.~M. Karp.
\newblock Algorithms for graph partitioning on the planted partition model.
\newblock \emph{Random Structures and Algorithms}, pages 116--140, 2001.

\bibitem[Dasgupta and Ghosh(2013)]{dasgupta2013crowdsourced}
A.~Dasgupta and A.~Ghosh.
\newblock Crowdsourced judgement elicitation with endogenous proficiency.
\newblock In \emph{World Wide Web (WWW)}, pages 319--330, 2013.

\bibitem[Decelle et~al.(2011)Decelle, Krzakala, Moore, and
  Zdeborov{\'a}]{decelle2011asymptotic}
A.~Decelle, F.~Krzakala, C.~Moore, and L.~Zdeborov{\'a}.
\newblock Asymptotic analysis of the stochastic block model for modular
  networks and its algorithmic applications.
\newblock \emph{Physical Review E}, 84\penalty0 (6), 2011.

\bibitem[Deng et~al.(2009)Deng, Dong, Socher, Li, Li, and
  Fei-Fei]{deng2009imagenet}
J.~Deng, W.~Dong, R.~Socher, L.~Li, K.~Li, and L.~Fei-Fei.
\newblock {I}mage{N}et: A large-scale hierarchical image database.
\newblock In \emph{Computer Vision and Pattern Recognition (CVPR)}, pages
  248--255, 2009.

\bibitem[Feige and Kilian(2001)]{feige2001heuristics}
U.~Feige and J.~Kilian.
\newblock Heuristics for semirandom graph problems.
\newblock \emph{Journal of Computer and System Sciences}, 63\penalty0
  (4):\penalty0 639--671, 2001.

\bibitem[Feige and Krauthgamer(2000)]{feige2000finding}
U.~Feige and R.~Krauthgamer.
\newblock Finding and certifying a large hidden clique in a semirandom graph.
\newblock \emph{Random Structures and Algorithms}, 16\penalty0 (2):\penalty0
  195--208, 2000.

\bibitem[Gu{\'e}don and Vershynin(2014)]{guedon2014community}
O.~Gu{\'e}don and R.~Vershynin.
\newblock Community detection in sparse networks via {G}rothendieck's
  inequality.
\newblock \emph{arXiv}, 2014.

\bibitem[Harmon(2004)]{harmon2004amazon}
A.~Harmon.
\newblock Amazon glitch unmasks war of reviewers.
\newblock \emph{New York Times}, 2004.

\bibitem[Holland et~al.(1983)Holland, Laskey, and
  Leinhardt]{holland1983stochastic}
P.~W. Holland, K.~B. Laskey, and S.~Leinhardt.
\newblock Stochastic blockmodels: Some first steps.
\newblock \emph{Social Networks}, 5:\penalty0 109--137, 1983.

\bibitem[Kamble et~al.(2015)Kamble, Shah, Marn, Parekh, and
  Ramachandran]{kamble2015truth}
V.~Kamble, N.~Shah, D.~Marn, A.~Parekh, and K.~Ramachandran.
\newblock Truth serums for massively crowdsourced evaluation tasks.
\newblock \emph{arXiv}, 2015.

\bibitem[Karger et~al.(2014)Karger, Oh, and Shah]{karger2014budget}
D.~R. Karger, S.~Oh, and D.~Shah.
\newblock Budget-optimal task allocation for reliable crowdsourcing systems.
\newblock \emph{Operations Research}, 62\penalty0 (1):\penalty0 1--24, 2014.

\bibitem[Krivelevich and Vilenchik(2006)]{krivelevich2006semirandom}
M.~Krivelevich and D.~Vilenchik.
\newblock Semirandom models as benchmarks for coloring algorithms.
\newblock In \emph{Meeting on Analytic Algorithmics and Combinatorics}, pages
  211--221, 2006.

\bibitem[Kulkarni et~al.(2015)Kulkarni, Koh, Huy, Chia, Papadopoulos, Cheng,
  Koller, and Klemmer]{kulkarni2015peer}
C.~Kulkarni, P.~W. Koh, H.~Huy, D.~Chia, K.~Papadopoulos, J.~Cheng, D.~Koller,
  and S.~R. Klemmer.
\newblock Peer and self assessment in massive online classes.
\newblock \emph{Design Thinking Research}, pages 131--168, 2015.

\bibitem[Le and Vershynin(2015)]{le2015concentration}
C.~M. Le and R.~Vershynin.
\newblock Concentration and regularization of random graphs.
\newblock \emph{arXiv}, 2015.

\bibitem[Makarychev et~al.(2012)Makarychev, Makarychev, and
  Vijayaraghavan]{makarychev2012approximation}
K.~Makarychev, Y.~Makarychev, and A.~Vijayaraghavan.
\newblock Approximation algorithms for semi-random partitioning problems.
\newblock In \emph{Symposium on Theory of Computing (STOC)}, pages 367--384,
  2012.

\bibitem[Makarychev et~al.(2015)Makarychev, Makarychev, and
  Vijayaraghavan]{makarychev2015learning}
K.~Makarychev, Y.~Makarychev, and A.~Vijayaraghavan.
\newblock Learning communities in the presence of errors.
\newblock \emph{arXiv}, 2015.

\bibitem[Mannor and Tsitsiklis(2004)]{mannor2004sample}
S.~Mannor and J.~N. Tsitsiklis.
\newblock The sample complexity of exploration in the multi-armed bandit
  problem.
\newblock \emph{Journal of Machine Learning Research (JMLR)}, 5:\penalty0
  623--648, 2004.

\bibitem[Massouli{\'e}(2014)]{massoulie2014community}
L.~Massouli{\'e}.
\newblock Community detection thresholds and the weak {R}amanujan property.
\newblock In \emph{Symposium on Theory of Computing (STOC)}, pages 694--703,
  2014.

\bibitem[Mayzlin et~al.(2012)Mayzlin, Dover, and
  Chevalier]{mayzlin2012promotional}
D.~Mayzlin, Y.~Dover, and J.~A. Chevalier.
\newblock Promotional reviews: An empirical investigation of online review
  manipulation.
\newblock Technical report, National Bureau of Economic Research, 2012.

\bibitem[Miller et~al.(2005)Miller, Resnick, and
  Zeckhauser]{miller2005eliciting}
N.~Miller, P.~Resnick, and R.~Zeckhauser.
\newblock Eliciting informative feedback: The peer-prediction method.
\newblock \emph{Management Science}, 51\penalty0 (9):\penalty0 1359--1373,
  2005.

\bibitem[Moitra et~al.(2015)Moitra, Perry, and Wein]{moitra2015robust}
A.~Moitra, W.~Perry, and A.~S. Wein.
\newblock How robust are reconstruction thresholds for community detection?
\newblock \emph{arXiv}, 2015.

\bibitem[Mossel et~al.(2012)Mossel, Neeman, and Sly]{mossel2012stochastic}
E.~Mossel, J.~Neeman, and A.~Sly.
\newblock Stochastic block models and reconstruction.
\newblock \emph{arXiv}, 2012.

\bibitem[Mossel et~al.(2013{\natexlab{a}})Mossel, Neeman, and
  Sly]{mossel2013belief}
E.~Mossel, J.~Neeman, and A.~Sly.
\newblock Belief propagation, robust reconstruction, and optimal recovery of
  block models.
\newblock \emph{arXiv}, 2013{\natexlab{a}}.

\bibitem[Mossel et~al.(2013{\natexlab{b}})Mossel, Neeman, and
  Sly]{mossel2013proof}
E.~Mossel, J.~Neeman, and A.~Sly.
\newblock A proof of the block model threshold conjecture.
\newblock \emph{arXiv}, 2013{\natexlab{b}}.

\bibitem[Mossel et~al.(2015)Mossel, Neeman, and Sly]{mossel2015consistency}
E.~Mossel, J.~Neeman, and A.~Sly.
\newblock Consistency thresholds for the planted bisection model.
\newblock In \emph{Symposium on Theory of Computing (STOC)}, pages 69--75,
  2015.

\bibitem[Piech et~al.(2013)Piech, Huang, Chen, Do, Ng, and
  Koller]{piech2013tuned}
C.~Piech, J.~Huang, Z.~Chen, C.~Do, A.~Ng, and D.~Koller.
\newblock Tuned models of peer assessment in {MOOC}s.
\newblock \emph{arXiv}, 2013.

\bibitem[Priedhorsky et~al.(2007)Priedhorsky, Chen, Lam, Panciera, Terveen, and
  Riedl]{priedhorsky2007creating}
R.~Priedhorsky, J.~Chen, S.~T.~K. Lam, K.~Panciera, L.~Terveen, and J.~Riedl.
\newblock Creating, destroying, and restoring value in {W}ikipedia.
\newblock In \emph{International {ACM} Conference on Supporting Group Work},
  pages 259--268, 2007.

\bibitem[Resnick and Sami(2007)]{resnick2007influence}
P.~Resnick and R.~Sami.
\newblock The influence limiter: provably manipulation-resistant recommender
  systems.
\newblock In \emph{ACM Conference on Recommender Systems}, pages 25--32, 2007.

\bibitem[Shah et~al.(2015)Shah, Zhou, and Peres]{shah2015approval}
N.~Shah, D.~Zhou, and Y.~Peres.
\newblock Approval voting and incentives in crowdsourcing.
\newblock In \emph{International Conference on Machine Learning (ICML)}, 2015.

\bibitem[Shah and Zhou(2015)]{shah2015double}
N.~B. Shah and D.~Zhou.
\newblock Double or nothing: Multiplicative incentive mechanisms for
  crowdsourcing.
\newblock In \emph{Advances in Neural Information Processing Systems (NIPS)},
  2015.

\bibitem[Shnayder et~al.(2016)Shnayder, Frongillo, Agarwal, and
  Parkes]{shnayder2016strong}
V.~Shnayder, R.~Frongillo, A.~Agarwal, and D.~C. Parkes.
\newblock Strong truthfulness in multi-task peer prediction, 2016.

\bibitem[Vuurens et~al.(2011)Vuurens, de~Vries, and Eickhoff]{vuurens2011spam}
J.~Vuurens, A.~P. de~Vries, and C.~Eickhoff.
\newblock How much spam can you take? {A}n analysis of crowdsourcing results to
  increase accuracy.
\newblock \emph{ACM SIGIR Workshop on Crowdsourcing for Information Retrieval},
  2011.

\bibitem[Zhang et~al.(2014)Zhang, Chen, Zhou, and
  Jordan]{zhang2014crowdsourcing}
Y.~Zhang, X.~Chen, D.~Zhou, and M.~I. Jordan.
\newblock Spectral methods meet {EM}: A provably optimal algorithm for
  crowdsourcing.
\newblock \emph{arXiv}, 2014.

\bibitem[Zhou et~al.(2015)Zhou, Liu, Platt, Meek, and
  Shah]{zhou2015regularized}
D.~Zhou, Q.~Liu, J.~C. Platt, C.~Meek, and N.~B. Shah.
\newblock Regularized minimax conditional entropy for crowdsourcing.
\newblock \emph{arXiv}, 2015.

\end{thebibliography}
}

\appendix
\section{Deferred Proofs}
\label{sec:proofs}

\subsection{Matrix Concentration Bound of \citet{le2015concentration}}
\label{sec:le-statement}

For ease of reference, here we state the matrix concentration bound 
from \citetm{le2015concentration}, which we make use of in the proofs below.
\begin{theorem}[Theorem 2.1 in~\citetm{le2015concentration}]\label{thm:le}
Given an $s \times s$ matrix $P$ with entries $P_{i,j} \in [0,1]$, and a random matrix $A$ with the properties that 1) each entry of $A$ is chosen independently, 2) $\E[A_{i,j}] = P_{i,j}$, and 3) $A_{i,j} \in [0,1]$, then for any $r \ge 1$, the following holds with probability at least $1-s^{-r}$:  let $d = s \cdot \max_{i,j} P_{i,j}$, and modify any subset of at most $10s/d$ rows and/or columns of $A$ by arbitrarily decreasing the value of nonzero elements of those rows or columns to form the matrix $A'$ with entries in $[0,1]$, then $$||A' - P||_{op} \le C r^{3/2}\left(\sqrt{d} + \sqrt{d'}\right),$$ where $d'$ is the maximum $\ell_2$ norm of any row or column of $A'$, and $C$ is an absolute constant.
\end{theorem}
Note: The proof of this theorem in~\citetm{le2015concentration} shows that the statement continues to hold in the slightly more general setting where the entries of $A$ are chosen independently according to random variables with bounded variance and sub-Gaussian tails, rather than just random variables restricted to the interval $[0,1]$.

\subsection{Details of Lipschitz Bound (Equation \ref{eq:lipschitz-usage})}
\label{sec:lipschitz-details}

The proof essentially consists of matching up each value $\ravg_j$, for 
$j \in T^*$, with a set of values $\ravg_{j'}$, $j' \geq j$, where the 
corresponding $\M_{i,j'}$ sum to $1$; we can then invoke the condition 
\eqref{eq:lipschitz}. Unfortunately, expressing this idea 
formally is a bit notationally cumbersome.

Before we start, we observe that the Lipschitz condition \eqref{eq:lipschitz} 
implies that, if $\ravg_j \geq \ravg_{j'}$, then 
$\ravg_j - \ravg_{j'} \leq L \cdot \p{\Aavg_{i,j} - \Aavg_{i,j'}} + \epsilon_0$. 
It is this form of \eqref{eq:lipschitz} that we will make use of below.

Now, let $I_j = \bI[j \in T^*]$, and without loss of generality suppose that 
the indices $j$ are such that $\ravg_1 \geq \ravg_2 \geq \cdots \geq \ravg_m$. 
For a vector $v \in [0,1]^m$, define
\begin{equation}
\label{eq:def-h}
h(\tau, v) \eqdef \inf\{j \mid \sum_{j' = 1}^j v_{j'} \geq \tau\},
\end{equation} 
where $h(\tau,v) = \infty$ if no such $j$ exists.
We observe that for any vector $v \in [0,1]^m$, we have 
\begin{equation}
\label{eq:sum-integral}
\sum_{j \in [m]} v_j\ravg_j = \int_{0}^{\infty} \ravg_{h(\tau; v)} d\tau,
\end{equation}
where we define $\ravg_{\infty} = 0$ (note that the integrand is therefore $0$ 
for any $\tau \geq \|v\|_1$). Hence, we have
\begin{align}
\sum_{j \in T^*} \ravg_j - \sum_{j \in [m]} \M_{i,j}\ravg_j 
 &= \sum_{j \in [m]} I_j\ravg_j - \sum_{j \in [m]} \M_{i,j}\ravg_j \\
 &= \int_{0}^{\beta m} \ravg_{h(\tau, I)} - \ravg_{h(\tau, \M_i)} d\tau \\
 &\stackrel{(i)}{\leq} \int_{0}^{\beta m} \left[L \cdot \p{\Aavg_{h(\tau, I)} - \Aavg_{h(\tau, \M_i)}} + \epsilon_0\right] d\tau \\
 &= L\cdot \p{\sum_{j \in [m]} I_j\Aavg_j - \sum_{j \in [m]} \M_{i,j}\Aavg_j} + \beta m\epsilon_0 \\
 &= L\cdot \p{\sum_{j \in T^*} \Aavg_j - \sum_{j \in [m]} \M_{i,j}\Aavg_j} + \beta m\epsilon_0,
\end{align}
which implies \eqref{eq:lipschitz-usage}. The key step is (i), 
which uses the fact that $h(\tau,I) \leq h(\tau, \M_i)$ (because $I$ is 
maximally concentrated on the left-most indices of $[m]$), and hence 
$\ravg_{h(\tau,I)} \geq \ravg_{h(\tau, \M_i)}$.

\subsection{Stability Under Noise (Proof of Lemma~\ref{lem:objective-bound})}
\label{sec:objective-bound-proof}

By H\"{o}lder's inequality, we have that $|\langle \A-\Aa, M \rangle| \leq \|\A-\Aa\|_{\op}\|M\|_*$.   We now leverage Theorem~\ref{thm:le} to bound $\|\A-\Aa\|_{\op}$.  To apply the theorem, first note that from the construction of $\A$ given in Algorithm~\ref{alg:create-A}, $\A$ can be constructed by first having the raters rate each item independently with probability $k/m$ to form matrix $\A^o$ and then removing ratings from the ``heavy'' rows (i.e. rows with more than $2k$ ratings), and ``heavy'' columns (i.e. columns with more than $2k$) ratings.  By standard Chernoff bounds, the probability that a given row or column will need to be ``pruned'' is at most $e^{-k/3} \le 2/k$, and hence from the independence of the rows, the probability that more than $5n/k$ rows are ``heavy'' is at most $e^{-2n/3k}$.  The probability that there are more than $5n/k$ heavy columns is identically bounded.

 Note that the expectation of the portion of $\A^o$ corresponding to the reliable raters is exactly the corresponding portion of matrix $\Aa,$ and with probability at least $1-2e^{-2n/3k}$, at most $10 n/k$ rows and/or columns of $\A^o$ are pruned to form $\A$.  Consider padding matrices $\A$ and $\Aa$ with zeros, to form the $n \times n$ matrices $\A'$ and $\Aa'$.   With probability $1-2e^{-2n/3k}$ the conditions of Theorem~\ref{thm:le} now apply to $\A'$ and $\Aa'$, with the parameters $d = \frac{n k}{m} \le k$, and $d' =  2k$.  Hence for any $r \ge 1$, with probability at least $1-2e^{-2n/3k} -n^{-r}$ $$\|\A-\Aa\|_{\op} = \|\A'-\Aa'\|_{\op} \le C r^{3/2} \sqrt{k},$$ for some absolute constant $C$.  

By assumption, $\|\M\|_* \le \frac{2}{\alpha \eps}\sqrt{\alpha \beta n m}$ and $\|\Mm\|_* \le \frac{2}{\alpha \eps}\sqrt{\alpha \beta n m}$.   Hence setting $r= \log (1/\delta),$ and $k \ge C' \log^3(\frac{1}{\delta}) \frac{m/n}{\eps^4 \alpha^3 \beta}$ for some absolute constant $C'$, we have that with probability at least $1-\delta$, we have $$|\langle \A-\Aa, \M \rangle|  \le \frac{1}{2} \eps \alpha \beta k n,$$ and $|\langle \A-\Aa, \Mm \rangle|$ is bounded identically.

To conclude, we have the following:
\begin{align}
\langle \Aa, \M \rangle  &\geq \langle \A, \M \rangle - \frac{1}{2}\epsilon \alpha \beta k n \\
 &\geq \langle \A, \Mm \rangle - \frac{1}{2} \epsilon \alpha \beta k n \text{ (since $\M$ is optimal for $\A$)} \\
 &\geq \langle \Aa, \Mm \rangle - \epsilon \alpha \beta k n,
\end{align}
which completes the proof.

\subsection{Bounding the Effect of Adversaries (Proof of Lemma~\ref{lem:subgradient})}
\label{sec:subgradient-proof}
In this section we prove Lemma~\ref{lem:subgradient}.
Let $\sP_0$ be the superset of $\sP$ where we have removed the 
nuclear norm constraint. By Lagrangian duality we 
know that there is some $\mu$ such that maximizing 
$\langle \Aa, M \rangle$ over $\sP \cap \{M_{\sC} = R_{\sC}\}$ 
is equivalent to maximizing $f_\mu(M) \eqdef \langle \Aa, M \rangle + \mu\p{\frac{2}{\epsilon\alpha}\sqrt{\alpha\beta nm} - \|M\|_*}$ over 
$\sP_0 \cap \{M_{\sC} = R_{\sC}\}$. 

We start by bounding $\mu$. We claim that 
$\mu \leq \epsilon k \sqrt{\alpha\beta n/m}$. 
To show this, we will first show that $\langle \Aa, M \rangle$ cannot get 
too large. Let $\Obs$ be the set of $(i,j)$ for which ratings are observed, 
and define the matrix $B'$ as 
$(B')_{ij} = \frac{k}{m} + \bI[(i,j) \in \Obs]\p{B_{ij} - 1}$; note 
that $(B-B')_{ij} = \bI[(i,j) \in \Obs] - \frac{k}{m}$.
For any $M \in \sP_0$, we have 
\begin{align}
\langle \Aa, M \rangle &\leq \langle B', M \rangle + \langle \Aa - B', M \rangle \\
 &\leq \beta kn + \|\Aa - B'\|_{\op}\|M\|_* \\
 &\stackrel{(i)}{\leq} \beta kn + \log(1/\delta)^{3/2} 2\sqrt{2k}\|M\|_* \\
 &\stackrel{(ii)}{\leq} k\p{\beta n +  \frac{\epsilon\sqrt{\alpha\beta n/m}}{2} \|M\|_*}.
\end{align}
In (i) we used the matrix concentration inequality of Theorem~\ref{thm:le}, in a similar manner as was used in our proof of Lemma~\ref{lem:objective-bound}.  Specifically, we consider padding $\Aa$ and $B'$ with zeros so as to make both into $n\times n$ matrices. Provided the total number of raters and items whose initial assignments are removed in the second and third steps of the rater/item assignment procedure (Algorithm~\ref{alg:create-A}) is bounded by $10n/k$, which occurs with probability at least $1-\delta/2$ given our choice of $k$, then Theorem~\ref{thm:le} applies with $r = \log (1/\delta)$, and $d$ and $d'$ bounded by $2k$, yielding an operator norm bound of $r^{3/2}(\sqrt{k} +\sqrt{2k}) \le \log(1/\delta)^{3/2} 2\sqrt{2k}$, that holds with probability $1-n^{-r} > 1-\delta/2$.  In (ii) we plug in our assumption $k = \Omega\p{\frac{\log(1/\delta)^3}{\alpha\beta\epsilon^2}\frac{m}{n}}$.

Now, suppose that we take $\mu_0 = \epsilon \sqrt{\alpha\beta n/m}k$ and optimize $\langle \Aa, M \rangle - \mu_0\|M\|_*$ over 
$\sP_0 \cap \{M_{\sC} = R_{\sC}\}$. By the above inequalities, 
we have $\langle \Aa, M \rangle - \mu_0\|M\|_* \leq \beta kn - \frac{\epsilon \sqrt{\alpha\beta n/m}k}{2}\|M\|_*$, 
and so any $M$ with $\|M\|_* > \frac{2}{\epsilon\alpha}\sqrt{\alpha\beta nm}$ 
cannot possibly be optimal, since the solution $M = 0$ would 
be better. Hence, $\mu_0$ is a large enough Lagrange multiplier to ensure that $M \in \sP$, and so 
$\mu \leq \mu_0 = \epsilon k\sqrt{\alpha\beta n/m}$, as claimed.

We next characterize the subgradient of $f_{\mu}$ at $M = \Mm$.
Define the projection matrix $P$ as
\[ P_{i,i'} = \left\{ \begin{array}{ccl} \frac{1}{|\sC|} & : & i, i' \in \sC \\ \delta_{i,i'} & : \text{else} \end{array} \right.. \]
Thus $PM = M$ if and only if all rows in $\sC$ are equal to each other.
In particular, $PM = M$ whenever $M_{\sC} = R_{\sC}$. Now, since $\Mm$ is the maximum 
of $f_{\mu}(M)$ over all $M \in \sP_0 \cap \{M_{\sC} = R_{\sC}\}$, there must be some 
$G \in \partial f_{\mu}(\Mm)$ such that $\langle G, M - \Mm \rangle \leq 0$ for all $M \in \sP_0 \cap \{M_{\sC} = R_{\sC}\}$. The following lemma says that without 
loss of generality we can assume that $PG = G$:
\begin{lemma}
\label{lem:subgradient-2}
Suppose that $G \in \partial f(\Mm)$ satisfies $\langle G, M - \Mm \rangle \leq 0$ 
for all $M \in \sP_0 \cap \{M_{\sC} = R_{\sC}\}$. 
Then, $PG$ satisfies the same property, and lies in $\partial f(\Mm)$ as well.
\end{lemma}
We can further note (by differentiating $f_{\mu}$) that 
$G = \Aa - \mu Z_0$, where $\|Z_0\|_{\op} \leq 1$\footnote{This is due to the 
more general result that, for any norm $\|\cdot\|$, the subgradient of $\|\cdot\|$ 
at any point has dual norm at most $1$.}. Then
$PG = P\Aa - \mu PZ_0 = \Aa - \mu PZ_0$. Let $r(M)$ denote the 
matrix where $M_{\sC}$ is replaced with $R_{\sC}$ (so $r(M) \in \sP_0 \cap \{R_{\sC} = M_{\sC}\}$ 
whenever $M \in \sP_0$). The rest of the proof is basically algebra; for any 
$M \in \sP$, we have
\begin{align}
\langle \Aa, M - \Mm \rangle &\stackrel{(i)}{\leq} f_{\mu}(M) - f_{\mu}(\Mm) \\
 &\stackrel{(ii)}{\leq} \langle \Aa - \mu PZ_0, M - \Mm \rangle \\
 &= \langle \Aa - \mu PZ_0, M - r(M) \rangle + \langle \Aa - \mu PZ_0, r(M) - \Mm \rangle \\
 &\stackrel{(iii)}{\leq} \langle \Aa - \mu PZ_0, M - r(M) \rangle \\
 &\stackrel{(iv)}{=} \langle \Aa_{\sC} - \mu (PZ_0)_{\sC}, M_{\sC} - r(M)_{\sC} \rangle \\
 &= \langle \Aa_{\sC} - \mu (PZ_0)_{\sC}, M_{\sC} - \Mm_{\sC} \rangle,
\end{align}
where (i) is by complementary slackness (either $\mu = 0$ or $\|\Mm\|_* = \frac{2}{\alpha\epsilon}\sqrt{\alpha\beta nm}$); 
(ii) is concavity of $f_{\mu}$, and the fact that $\Aa - \mu PZ_0$ is a subgradient;  
(iii) is the property from Lemma~\ref{lem:subgradient-2} ($\langle \Aa - \mu PZ_0, r(M) - \Mm \rangle \leq 0$ since 
$r(M) \in \sP_0$); and (iv) is because $M$ and $r(M)$ only differ on $\good$.

To finish, we will take $Z = \mu (PZ_0)_{\sC}$. We note that
$\|Z\|_{\op} = \|\mu (PZ_0)_{\sC}\|_{\op} \leq \mu \|PZ_0\|_{\op} \leq \mu \|Z_0\|_{\op} \leq \mu$.
Moreover, $Z$ has rank $1$ and so $\|Z\|_F = \|Z\|_{\op} \leq \mu \leq \epsilon k\sqrt{\alpha\beta n/m}$, as was to be shown.

\paragraph{Proof of Lemma~\ref{lem:subgradient-2}.}
First, since $PM = M$ for all $M \in \sP_0 \cap \{M_{\sC} = R_{\sC}\}$, and $PM^* = M^*$, 
we have $\langle PG, M-M^* \rangle = \langle G, P(M-M^*) \rangle = \langle G, M-M^* \rangle \leq 0$. 
We thus only need to show that $PG$ is still a subgradient of $f_{\mu}$. Indeed, we have (for arbitrary $M$)
\begin{align}
\langle PG, M-M^* \rangle &= \langle G, PM - M^* \rangle \\
 &\stackrel{(i)}{\geq} f_{\mu}(PM) - f_{\mu}(M^*) \\
 &= \langle \Aa, PM \rangle - \mu \|PM\|_* - f_{\mu}(M^*) \\
 &= \langle \Aa, M \rangle - \mu \|PM\|_* - f_{\mu}(M^*) \\
 &\stackrel{(ii)}{\geq} \langle \Aa, M \rangle - \mu \|M\|_* - f_{\mu}(M^*) \\
 &= f_{\mu}(M) - f_{\mu}(M^*),
\end{align}
where (i) is because $G \in \partial f_{\mu}(M^*)$, and (ii) is because projecting 
decreases the nuclear norm. Since the inequality $\langle PG, M-M^* \rangle \geq f_{\mu}(M) - f_{\mu}(M^*)$ 
is the defining property for $PG$ to lie in $\partial f_{\mu}(M^*)$, the proof is complete.

\subsection{Proof of Proposition~\ref{prop:recover-M}}
\label{sec:recover-M-proof}
In this section, we will prove Proposition~\ref{prop:recover-M} 
from Lemmas~\ref{lem:objective-bound} and \ref{lem:subgradient}.
We start by plugging in $\M$ for $M$ in Lemma~\ref{lem:subgradient}. This yields
$\langle \Aa_{\good} - Z_{\good}, \Mm_{\good} - \M_{\good} \rangle \leq \langle \Aa, \Mm - \M \rangle \leq \epsilon \alpha\beta kn$
by Lemma~\ref{lem:objective-bound}.
On the other hand, we have 
\begin{align}
|\langle Z_{\good}, \Mm_{\good} - \M_{\good} \rangle| &\leq \|Z_{\good}\|_F\|\Mm_{\good} - \M_{\good}\|_F \\
 &\leq \epsilon\sqrt{\alpha\beta nk/m} \sqrt{\|\Mm_{\good} - \M_{\good}\|_{1}\|\Mm_{\good} - \M_{\good}\|_{\infty}} \\
 &\leq \epsilon\sqrt{\alpha\beta nk/m} \sqrt{2\alpha\beta mn} = \sqrt{2}\epsilon\alpha\beta kn.
\end{align}
Putting these together, we obtain
$\langle \Aa_{\good}, \Mm_{\good} - \M_{\good} \rangle \leq (1+\sqrt{2})\epsilon \alpha\beta kn$.
Expanding $\langle \Aa_{\good}, \Mm_{\good} - \M_{\good} \rangle$ as 
$\frac{k}{m}\sum_{i \in \good}\p{\sum_{j \in [m]} (R_{ij} - \M_{ij})\Aavg_{ij}}$,
we obtain 
\[ \frac{1}{|\good|}\frac{1}{\beta m}\sum_{i \in \good}\sum_{j \in [m]} (T_j^*-\M_{ij})\Aavg_{ij} \leq (1+\sqrt{2})\epsilon. \]
Scaling $\epsilon$ by a factor of $1+\sqrt{2}$ yields the desired result.

\subsection{Concentration Bounds for $\robs$ (Proof of Lemma~\ref{lem:robs-rtrue})}
\label{sec:concentration-proof}
We start by stating a lemma which will be useful both here and later:
\begin{lemma}
\label{lem:chernoff}
Let $M \in [0,1]^{n \times m}$ be a matrix of random variables 
such that $\|M_i\|_2^2 \leq \beta m$ for all rows $i \in [n]$. 
Define the deviation $D_i \eqdef \sum_{j=1}^m \M_{ij}(\robs_j - \frac{k_0}{m}\ravg_j)$. 
Then, for $k_0 \geq \frac{3\log(2n/v\delta)}{\min(\epsilon,\epsilon^2)}$, 
with probability $1-\delta$, we have 
$\left|\frac{1}{|V|} \sum_{i \in V} D_{i}\right| \leq \epsilon \beta k_0$ for all 
sets $V \subseteq [n]$ with $|V| \geq v$.
\end{lemma}
Given Lemma~\ref{lem:chernoff}, the rest of the proof is fairly straightforward.
Noting that $\epsilon \leq 1$ and applying this conclusion for 
$v = \alpha n$, and $k_0 \geq \frac{3 \cdot 8^2\log(2/\alpha\delta)}{\epsilon^2}$, 
we see that
\begin{align}
\frac{1}{\alpha n} \sum_{i \in \goodapprox} \sum_{j \in [m]} \M_{ij} \ravg_{ij} &\geq \frac{1}{\alpha n}\frac{m}{k_0} \sum_{i \in \goodapprox} \sum_{j \in [m]} \M_{ij} \robs_{ij} - \frac{\epsilon}{8} \beta m \\
 &\geq \frac{1}{|\good|}\frac{m}{k_0} \sum_{i \in \good} \sum_{j \in [m]} \M_{ij} \robs_{ij} - \frac{\epsilon}{8} \beta m \\
 &\geq \frac{1}{|\good|}\sum_{i \in \good} \sum_{j \in [m]} \M_{ij} \ravg_{ij} - \frac{\epsilon}{4} \beta m,
\end{align}
as was to be shown.

\paragraph{Proof of Lemma~\ref{lem:chernoff}.}
Define the cumulant function $c_i(\lambda) \eqdef \log(\bE_{\robs}[\exp(\lambda D_i)])$. We have
\begin{align}
c_i(\lambda) &=\log(\bE_{\robs}[\exp(\lambda \sum_j \M_{ij}(\robs_j - (k_0/m)\ravg_j))]) \\
 &=\sum_j \log(\bE_{\robs}[\exp(\lambda \M_{ij}(\robs_j - (k_0/m)\ravg_j))]) \\
 &\stackrel{(i)}{\leq} \sum_j (e^{\lambda} - \lambda - 1)\M_{ij}^2\Var[\robs_j] \\
 &\leq (e^{\lambda} - \lambda - 1) \sum_j \M_{ij}^2 \frac{k_0}{m} \\
 &\leq (e^{\lambda}-\lambda-1)\beta k_0,
\end{align}
where (i) is Bennet's inequality.

We also consider the cumulant function for the maximum average deviation over 
possible sets $V$:
\begin{equation}
\label{eq:def-C}
C_v(\lambda) \eqdef \log\p{\bE_{\robs}\left[ \max_{|V| \geq v} \exp\p{\frac{\lambda}{|V|} \sum_{i \in V} D_{i}}\right]}.
\end{equation}
To bound $C_v(\lambda)$, we use the power mean inequality
\begin{align}
\max_{|V| \geq v} \exp\p{\frac{\lambda}{|V|} \sum_{i \in V} D_{i}} 
 &\leq \max_{|V| \geq v} \frac{1}{|V|} \sum_{i \in V} \exp\p{\lambda D_{i}} \\
 &\leq \max_{|V| \geq v} \frac{1}{|V|} \sum_{i=1}^n \exp\p{\lambda D_i} \\
 &\leq \frac{1}{v} \sum_{i=1}^n \exp\p{\lambda D_i}.
\end{align}
Therefore, 
\begin{align}
C_v(\lambda) &= \log\p{\bE_{\robs}\left[ \max_{|V| \geq v} \exp\p{\frac{\lambda}{|V|} \sum_{i \in V} D_{i}}\right]} \\
 &\leq \log\p{\bE_{\robs}\left[\frac{1}{v} \sum_{i=1}^n \exp\p{\lambda D_i}\right]} \\
 &\leq \log\p{\frac{n}{v}\exp\p{(e^{\lambda} - \lambda - 1)\beta k_0}} \\
 &= \log(n/v) + (e^{\lambda} - \lambda - 1)\beta k_0.
\end{align}
By applying a standard Chernoff bound argument to $C_v(\lambda)$, we obtain
\begin{equation}
\label{eq:chernoff-conclusion}
\bP\left[\max_{|V| \geq v} \left|\frac{1}{|V|} \sum_{i \in V} D_{i}\right| \geq \epsilon \beta k_0\right] \leq \frac{2n}{v}\exp\p{-\frac{\beta k_0}{3}\min(\epsilon,\epsilon^2)}.
\end{equation}
In particular, for
$k_0 \geq \frac{3\log(2n/v\delta)}{\beta\min(\epsilon,\epsilon^2)}$, 
we have with probability $1-\delta$ that 
$\left|\frac{1}{|V|} \sum_{i \in V} D_{i}\right| \leq \epsilon \beta k_0$ 
for all sets $V \subseteq [n]$ with $|V| \geq v$, as was to be shown.

\subsection{Correctness of Randomized Rounding (Proof of Lemma~\ref{lem:rounding})}
\label{sec:rounding-proof}

Our goal is to show that the output of 
Algorithm~\ref{alg:round} satisfies $\bE[T] = T_0$. First, observe that 
since $(T_0)_j \leq 1$ for all $j$, each interval $[s_{j-1},s_j)$ has length 
at most $1$, and so the for loop over $z$ never picks the same index $j$ 
twice. Moreover, the probability that $j$ is included in $T_0$ is exactly 
$s_j - s_{j-1} = (T_0)_j$. The result follows by linearity of expectation.

\subsection{Correctness of Algorithm~\ref{alg:recover-T} (Proof of Proposition~\ref{prop:recover-T})}
\label{sec:recover-T-proof}

First, we claim that with probability $1-\delta$, we will invoke 
\texttt{RandomizedRound} at most $\frac{4\log(1/\delta)}{\epsilon\beta}$ times. 
To see this, note that $\bE[\langle T, \robs' \rangle] = \langle T_0, \robs' \rangle$, and 
$\langle T, \robs' \rangle \in [0, k_0]$ almost surely. 
By Markov's inequality, the probability that 
$\langle T, \robs' \rangle < \langle T_0, \robs' \rangle - \frac{\epsilon}{4}\beta k_0$ is at most $\frac{k_0 - \langle T_0, \robs' \rangle}{k_0 - \langle T_0, \robs' \rangle + (\epsilon/4)\beta k_0}$. We can assume that 
$\langle T_0, \robs' \rangle \geq (\epsilon/4)\beta k_0$ (since otherwise 
we accept $T$ with probability $1$), in which case the preceding expression 
is bounded by 
$\frac{k_0 - (\epsilon/4)\beta k_0}{k_0} = 1 - \frac{\epsilon}{4}\beta$. 
Therefore, the probability of accepting $T$ in any given iteration of the while 
loop is at least $\frac{\epsilon}{4}\beta$, and so the probability of 
accepting at least once in $\frac{4\log(1/\delta)}{\epsilon\beta}$ iterations is 
indeed at least $1 - \delta$.

Next, for $k_0 \geq \Omega\p{\frac{\log(2/\epsilon\beta\delta)}{\beta \epsilon^2}}$, 
we can make the probability that $|\langle T, \robs - \frac{k_0}{m}\ravg \rangle| \geq \frac{\epsilon}{4} \beta k_0$ be at most $\frac{\delta\epsilon\beta}{4\log(1/\delta)+1}$ (this follows from a standard Chernoff argument which we omit; 
Lemma~\ref{lem:chernoff} contains a superset of the necessary ideas). 
Therefore, by union bounding over the $\frac{4\log(1/\delta)}{\epsilon\beta}$ 
possible $T$ as well as $T_0$, with probability $1-2\delta$ we have $|\langle T, \robs - \frac{k_0}{m} \ravg \rangle| \leq \frac{\epsilon}{4}\beta k_0$ for whichever $T$ we end up accepting, as well as for $T = T_0$.

Consequently, we have
\begin{align}
\langle T, \ravg \rangle &\geq \frac{m}{k_0} \langle T, \robs' \rangle - \frac{\epsilon}{4}\beta m \\
 &\geq \frac{m}{k_0} \langle T_0, \robs' \rangle - \frac{2\epsilon}{4}\beta m \\
 &\geq \langle T_0, \ravg \rangle - \frac{3\epsilon}{4}\beta m \\
 &\geq \langle \frac{1}{|\good|} \sum_{i \in \good} \M_i,\, \ravg \rangle - \epsilon\beta m,
\end{align}
where the final step is Lemma~\ref{lem:robs-rtrue}.
By scaling down the failure probability $\delta$ by a constant 
(to account for the probability of failure at each step of the above argument), 
Proposition~\ref{prop:recover-T} follows.

\subsection{Proof of Theorem~\ref{thm:main}}
\label{thm:main-proof}

By Proposition~\ref{prop:recover-M}, for $k = \Omega\p{\frac{1}{\beta\alpha^3\epsilon^4}\max\p{1,\frac{m}{n}}}$, 
we can recover a matrix $\M$ satisfying 
$\frac{1}{|\good|}\frac{1}{\beta m}\sum_{i \in \good}\sum_{j \in [m]} (\Mm_{ij}-T^*_{j})\Aavg_{ij} \leq \epsilon$, 
and hence by \eqref{eq:lipschitz-usage} $\M$ also satisfies
$\frac{1}{|\good|}\frac{1}{\beta m}\sum_{i \in \good}\sum_{j \in [m]} (\Mm_{ij}-T^*_{j})\ravg_{j} \leq L \cdot \epsilon + \epsilon_0$.

By Proposition~\ref{prop:recover-T}, we then recover a set $T$ satisfying
\begin{align}
\frac{1}{\beta m}\sum_{j \in T}\ravg_j &\geq \p{\frac{1}{\beta m}\frac{1}{|\good|}\sum_{i \in \good}\sum_{j \in [m]} \M_{ij}\ravg_j} - \epsilon \\
 &\geq \p{\frac{1}{\beta m}\frac{1}{|\good|}\sum_{i \in \good}\sum_{j \in [m]} T^*_{j}\ravg_j} - [(L+1)\cdot\epsilon+\epsilon_0] \\
 &= \p{\frac{1}{\beta m}\sum_{j \in T^*} \ravg_j} - [(L+1)\cdot\epsilon+\epsilon_0],
\end{align}
as was to be shown.

\section{Examples of Adversarial Behavior}
\label{sec:adversary-examples}

In this section, in order to provide some intuition we show 
two possible attacks that adversaries could employ to make it 
hard for us to recover the good items. The first attack creates a 
symmetric situation, whereby there are $\frac{1}{\alpha}$ indistinguishable 
sets of potentially good items, and we are therefore forced to consider 
each set before we can find out which one is actually good. The 
second attack demonstrates the necessity of constraining each row 
to have a fixed sum, by showing that adversaries that are allowed to 
create very dense rows can have disproportionate influence on nuclear norm-based 
recovery algorithms

\subsection{Necessity of Nuclear Norm Scaling}

Suppose for simplicity that $\alpha = \beta$ and $n = m$. Let $J$ be the 
$\alpha n \times \alpha n$ all-ones matrix, and suppose that the full 
rating matrix $A$ has a block structure:
\begin{equation}
A^* = \left[ \begin{array}{cccc} J & (1-\epsilon)J & \cdots & (1-\epsilon)J \\ (1-\epsilon)J & J & \cdots & (1-\epsilon)J \\ \vdots & \vdots & \ddots & \vdots \\ (1-\epsilon)J & (1-\epsilon)J & \cdots & J \end{array} \right]
\end{equation}
In other words, both the items and raters are partitioned into $\frac{1}{\alpha}$ 
blocks, each of size $\alpha n$. A rater assigns a rating of $1$ to 
everything in their corresponding block, and a rating of $1-\epsilon$ to 
everything outside of their block. Thus, there are $\frac{1}{\alpha}$ completely 
symmetric blocks, only one of which corresponds to the good raters. Since we do 
not know which of these blocks is actually good, we need to include them all 
in our solution $M^*$. Therefore, $M^*$ should be
\begin{equation}
M^* = \left[ \begin{array}{cccc} J & 0 & \cdots & 0 \\ 0 & J & \cdots & 0 \\ \vdots & \vdots & \ddots & \vdots \\ 0 & 0 & \cdots & J \end{array} \right]
\end{equation}
Note however that in this case, $\|M^*\|_* = n$, while 
$\sqrt{\alpha\beta nm} = \sqrt{\alpha^2n^2} = \alpha n$. We therefore need the 
nuclear norm constraint in \eqref{eq:optimization-noisy} to be at least 
$\frac{1}{\alpha}$ times larger than $\sqrt{\alpha\beta nm}$ in order to capture 
the solution $M^*$ above.

It is not obvious to us that the additional $\frac{2}{\epsilon}$ factor appearing 
in \eqref{eq:optimization-noisy} is actually necessary, but it was needed in our 
analysis in order to bound the impact of adversaries.

\subsection{Necessity of Row Normalization}

Suppose that we did not include the row-normalization constraint 
$\sum_j \M_{ij} \leq \beta m$ in \eqref{eq:optimization-noisy}. For instance, 
this might happen if, instead of seeking all items of quality above a given 
quantile, we sought all items with quality above a given \emph{threshold} (say, 
whose quality was great than $\frac{1}{2}$). In this case we might pose the 
optimization problem
\begin{align}
\label{eq:optimization-naive}
\text{maximize } &\langle \A-\tfrac{1}{2}J_{n,m}, M \rangle, \\
\notag \text{ subject to } &0 \leq M_{ij} \leq 1 \,\,\, \forall i,j, \\
\notag                     &\|M\|_* \leq \frac{2}{\alpha\epsilon}\sqrt{\alpha\beta nm},
\end{align}
where $J_{n,m}$ is the $n \times m$ all-ones matrix. There are several reasons not 
to do this (for instance, focusing on quality thresholds rather than quantile 
thresholds loses the robustness to monotonic transformations that our method 
enjoys). In this section, we will focus on the particular issue that 
\eqref{eq:optimization-naive} is \emph{less robust to adversaries} than 
\eqref{eq:optimization-noisy}.

Concretely, we will suppose that the adversaries are split into 
$\frac{1}{3\beta}\p{\frac{1}{\alpha}-1}$ blocks of size $3\alpha\beta n$, 
each of which rates a random subset of 
$\frac{m}{2}$ items positively and the rest negatively. So for instance 
the matrix $A^*$ might look like (with $\alpha=\frac{2}{5}, \beta=\frac{1}{6}, n=10, m=12$):
\begin{equation}
\label{eq:row-construction}
\def\arraystretch{1.5}
A^* = \begin{blockarray}{c[cccccccccccc]}
\multirow{4}{*}{\rotatebox[origin=c]{90}{\textrm{good}}} 
 & 1 & 1 & 0 & 0 & 0 & 0 & 0 & 0 & 0 & 0 & 0 & 0 \topstrut \\
&  1 & 1 & 0 & 0 & 0 & 0 & 0 & 0 & 0 & 0 & 0 & 0 \\
&  1 & 1 & 0 & 0 & 0 & 0 & 0 & 0 & 0 & 0 & 0 & 0 \\
&  1 & 1 & 0 & 0 & 0 & 0 & 0 & 0 & 0 & 0 & 0 & 0 \\
\cline{2-13}
\multirow{2}{*}{\rotatebox[origin=c]{90}{\textrm{bad $1$}}}
 & 0 & 1 & 0 & 0 & 1 & 0 & 0 & 0 & 1 & 1 & 1 & 1 \\
&  0 & 1 & 0 & 0 & 1 & 0 & 0 & 0 & 1 & 1 & 1 & 1 \\
\cline{2-13}
\multirow{2}{*}{\rotatebox[origin=c]{90}{\textrm{bad $2$}}}
 & 0 & 1 & 1 & 0 & 1 & 1 & 0 & 1 & 0 & 0 & 1 & 0 \\
&  0 & 1 & 1 & 0 & 1 & 1 & 0 & 1 & 0 & 0 & 1 & 0 \\
\cline{2-13}
\multirow{2}{*}{\rotatebox[origin=c]{90}{\textrm{bad $3$}}}
 & 1 & 0 & 1 & 1 & 0 & 0 & 1 & 0 & 0 & 1 & 0 & 1 \\
&  1 & 0 & 1 & 1 & 0 & 0 & 1 & 0 & 0 & 1 & 0 & 1 \botstrut \\
\end{blockarray}
\end{equation}
The nuclear norm of each individual bad block is 
$\sqrt{\frac{3}{2}\alpha\beta nm}$, and because the blocks are 
chosen independently of each other, the nuclear norm will be approximately 
additive across blocks. In addition, including a given bad block 
increases $\langle \A - \frac{1}{2}J, M \rangle$ by $\frac{3}{4}\alpha\beta nm$.
In contrast, including the good block increases the nuclear norm by 
$\sqrt{\alpha\beta nm}$ and only increases the objective by $\frac{1}{2}\alpha\beta nm$. The bad blocks therefore all give more ``bang for the buck'' in terms of 
how much they increase the objective vs. how much much they increase the nuclear 
norm, so we will add them before the good block.

To accomodate all these bad 
blocks, we need to allow $\|M\|_*$ to be at least roughly 
$\frac{1}{3\beta}\p{\frac{1}{\alpha}-1} \times \sqrt{\frac{3}{2}\alpha\beta nm} 
= \Omega\p{\frac{1}{\alpha\beta}\sqrt{\alpha\beta nm}}$, which is adds an extra 
factor of $\frac{1}{\beta}$ relative to when we constrain the column sum. 
The issue can be seen in the above construction in \eqref{eq:row-construction}: 
if we do not normalize the rows, then the rows controlled by adversaries can 
exert disproportionate influence (up to a factor of $\frac{1}{\beta}$) by 
creating columns that are much denser than those of the reliable raters.


\end{document}